\documentclass[traditabstract]{aa1} 
\usepackage{graphicx}
\usepackage{txfonts}
\usepackage{natbib}
\usepackage{epstopdf}
\usepackage{longtable}
\usepackage{color}

\newcommand{\wn}{cm$^{-1}$}

\begin{document}

\title{
Using radio astronomical receivers for molecular spectroscopic characterization in
astrochemical laboratory simulations: A proof of concept
}

\author{
I. Tanarro\inst{1},
B. Alem\'an\inst{2},
P. de Vicente\inst{3},
J.D. Gallego\inst{3},
J.R. Pardo\inst{2},
G. Santoro\inst{4},
K. Lauwaet\inst{4},
F. Tercero\inst{3},
A. D\'{\i}az-Pulido\inst{3},
E. Moreno\inst{2},
M. Ag\'undez\inst{2},
J.R. Goicoechea\inst{2},
J.M. Sobrado\inst{5},
J.A. L\'opez\inst{3},
L. Mart\'{\i}nez\inst{4},
J.L. Dom\'enech\inst{1},
V.J. Herrero\inst{1},
J.M. Hern\'andez\inst{3},
R.J. Pel\'aez\inst{1},
J.A. L\'opez-P\'erez\inst{3},
J. G\'omez-Gonz\'alez\inst{3},
J.L. Alonso\inst{6},
E. Jim\'enez\inst{7},
D. Teyssier\inst{8},
K. Makasheva\inst{9},
M. Castellanos\inst{2},
C. Joblin\inst{10,11},
J.A. Mart\'{\i}n-Gago\inst{4},
J. Cernicharo\inst{2}\thanks{corresponding author. email: jose.cernicharo@csic.es}
}

\institute{
IEM. CSIC. Instituto de Estructura de la Materia. Molecular Physics Department. 
C/Serrano 123, 28006 Madrid, Spain.
\and
ICMM. CSIC. Molecular Astrophysics Group. C/ Sor Juana In\'es de la Cruz 3. Cantoblanco,
28049 Madrid. Spain
\and
Centro Nacional de Tecnolog\'{\i}as Radioastron\'omicas y Aplicaciones Geoespaciales (CNTRAG), Observatorio de Yebes (IGN), Spain
\and
ICMM. CSIC. Materials Science Factory. Structure of Nanoscopic Systems Group, ESISNA. C/ Sor Juana In\'es de la Cruz 3. Cantoblanco, 28049 Madrid. Spain
\and
Centro de Astrobiolog\'{\i}a, (CAB-CSIC/INTA). Carretera Torrej\'on a Ajalvir km 4, Torrej\'on de Ardoz 28850 (Madrid), Spain
\and
Grupo de Espectroscop\'{\i}a Molecular (GEM), Edificio Quifima, \'Area de Qu\'{\i}mica-F\'{\i}sica, Laboratorios de 
Espectroscop\'{\i}a y Bioespectroscop\'{\i}a, Parque Cient\'{\i}fico UVa, Unidad Asociada CSIC, Universidad de Valladolid, 
47011 Valladolid, Spain 
\and
Departamento de Qu\'imica F\'isica, Facultad de Ciencias y Tecnolog\'{\i}as Qu\'{\i}micas, 
Universidad de Castilla-La Mancha, Avda. Camilo Jos\'e Cela 1B, E-13071, Ciudad Real, Spain
\and
European Space Astronomy Centre, ESA, PO Box 78, 28691 Villanueva de la Cañada, Madrid, Spain
\and
LAPLACE (Laboratoire Plasma et Conversion d\'Energie); Universit\'e de Toulouse; CNRS, UPS, INPT; 
118 route de Narbonne, F-31062 Toulouse cedex 9, France
\and
Universit\'e de Toulouse, UPS-OMS, IRAP, 31000 Toulouse, France
\and
CNRS, IRAP, 9 Av. Colonel Roche, BP 44346, 31028 Toulouse Cedex 4, France
}

\date{Received  21 September 2017; accepted 21 September 2017}

\abstract{We present a proof of concept on the coupling of
radio astronomical receivers and spectrometers 
with chemical reactors
and the performances of the resulting setup for spectroscopy and chemical 
simulations in laboratory astrophysics. 
Several experiments including cold plasma generation and UV photochemistry were 
performed in a 40\,cm long gas cell placed in the beam path of the Aries 40\,m radio telescope receivers 
operating in the 41-49 GHz frequency range interfaced with fast Fourier transform spectrometers providing
2 GHz bandwidth and 38 kHz resolution.
 
The impedance matching of the cell windows has been studied using different materials. 
The choice of the material and its thickness was critical to obtain a sensitivity
identical to that of standard radio astronomical observations.

Spectroscopic signals arising from very low partial pressures of CH$_3$OH, CH$_3$CH$_2$OH, HCOOH, OCS,
CS, SO$_2$ ($<$10$^{-3}$ mbar) were detected in a few seconds. 
Fast data acquisition was achieved allowing for kinetic measurements in fragmentation experiments using 
electron impact or UV irradiation. Time evolution of chemical reactions involving OCS, O$_2$ and CS$_2$ 
was also observed demonstrating that reactive species, such as CS, can be maintained
with high abundance in the gas phase during these experiments.}


\titlerunning{A gas cell for Astrochemical simulations}

\authorrunning{Tanarro et al.}

\maketitle

\section{Introduction}
Molecular spectroscopy is the basic tool used to study the chemical, physical,
and dynamical evolution of {molecular clouds,} where stars and their planets form. 
Radio astronomers use large single dishes or interferometers equipped with
very sensitive receivers to
detect the thermal emission of the molecules present in these objects. Although volume
densities in molecular clouds, mostly corresponding to H$_2$, are rather low, 
$\simeq$10$^2$-10$^5$ cm$^{-3}$, the large size of these objects (typically several parsecs
$\simeq$10$^{19}$ cm), allows column densities to be rather high, $\sim$10$^{21}$-10$^{24}$ cm$^{-2}$.   
For a molecular cloud with a column density of 10$^{22}$ H$_2$ molecules\,cm$^{-2}$
(for example a prestellar core such as TMC1), the column densities of detected polar species  
vary from $\sim$10$^{11}$, for the less abundant ones, to $\simeq$10$^{18}$ cm$^{-2}$ for CO.
Using large single-dish telescopes (such as the 30m IRAM radio telescope) or interferometers (ALMA, NOEMA), the detection
of the weakest and narrowest molecular lines in cold dark clouds, requiring $\sim$20-30 kHz
spectral resolution, typically requires a few hours of observing time. On the other hand, the
broadest and most intense molecular lines in space are detected in a matter of minutes using
frequency resolutions of the order of 1 MHz with the same facilities.

In order to simulate one of these molecular clouds here on Earth we would have to compress and confine
the whole column of gas that a telescope observes within its beam into a gas cell which typically would 
have a length of the order of one meter. Column densities of the less abundant, but detected, molecules in space 
(N$\simeq$10$^{11}$ cm$^{-2}$) would translate into partial pressures inside such a chamber in the range
$\sim$10$^{-8}$ mbar (at T$_K\sim$10 K) to $\sim$5\,10$^{-6}$ mbar (at T$_K$$\sim$300 K). The idea behind
this work is that an astronomical receiver placed in front of such a gas cell would detect lines of these
low-abundance species in minutes { to} hours 

\begin{center}
\begin{figure}
    \label{fg:setup-1}
\includegraphics[angle=0,scale=0.42]{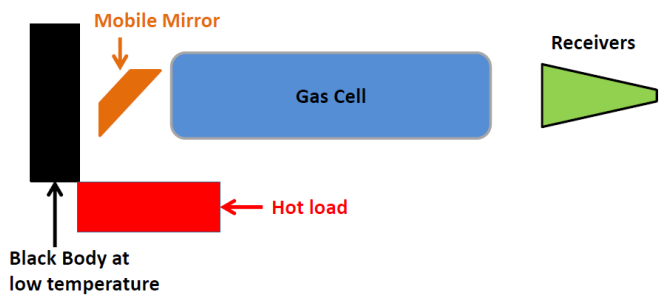}
\caption{Basic design concept of the gas cell setup.}
\label{fig:design}
\end{figure}
\end{center}

Radio astronomical receivers nowadays permit broad instantaneous
frequency coverage (several GHz) that, in turn, can be fully covered with Fast Fourier Transform
Spectrometers (FFTS) providing spectral resolutions as low as a few kHz. 
Standard techniques for millimeter laboratory spectroscopy use different approaches to provide
high sensitivity, broad band coverage, and spectral resolution such as the Chirped pulse Fourier 
Transform microwave (CHIRP) spectroscopy \citep{Brown2008,Dian2008}, extended to the millimeter region
as presented in \citet{Park2011}. However, sensitivity decreases considerably with increasing frequency.
 Chirped pulse spectroscopy has been applied by various groups to study kinetics and gas
compositions \citep{Abeysekera2014,Gerecht2011}.
Although the CHIRP pulse is not a passive method it has been 
shown that the initial population difference does not change in the linear fast passage regime 
for weak coupling (weak pulse limit) normally 
assumed. However, a large change in the population difference within the rapid adiabatic passage 
regime is observed for the strong 
coupling regime (strong pulse or high dipole moment). Here, stepwise multi-resonance effects can alter the relative 
intensities of the observed rotational 
transitions. In fact, the intensities differ significantly between measurements with chirp-up and chirp-down 
\citep{Schmitz2012},
which could cause serious difficulties in the
estimation of the abundance of the gases in the chamber.
Partial pressures, however, can be easily obtained from
the measurement of the molecular thermal emission using a radio astronomical receiver, just as radio astronomers do
when observing interstellar clouds.

\begin{figure}
\label{fg:setup-2}
\centering
\includegraphics[angle=0,width=9.0cm]{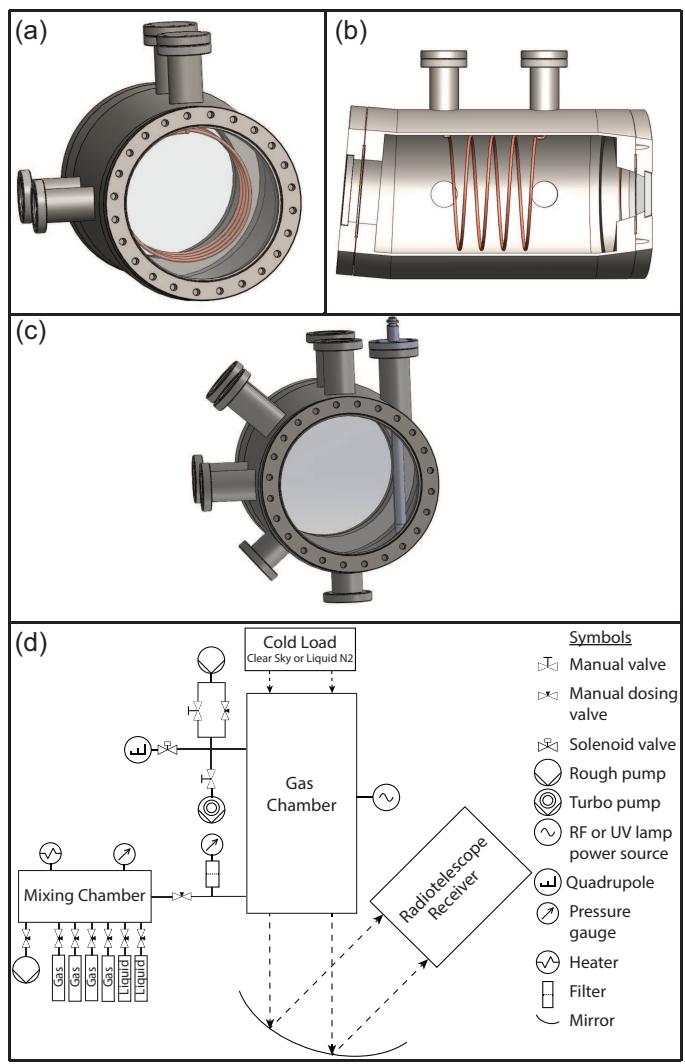}
\caption{Design of the gas cell prototype (a), its configuration as plasma reactor (b) and with the UV lamp (c), and
experimental set-up showing the technical scheme of the gas mixing, mass spectrometry,
and gas phase molecular detection (d).} 
\label{FigSetup}
\end{figure}

Within the { European Research Council} 21 September 2017
 Synergy project NANOCOSMOS we proposed 
using radio astronomical receivers as detectors for molecular spectroscopy and chemical reactivity 
experiments in a gas cell. The aim
of this work is therefore to present our first proof-of-concept and to shed light on the main issues related to the
technique. Different experiments have been performed in the past based on similar ideas 
\citep{Huberts2006,Ren2010,Neumaier2014,Wehres2016}, in particular for 
the SWAS satellite for which an OCS gas cell was used \citep{Tolls2004}, 
the Herschel/HIFI receivers
\citep{Higgins2010,Higgins2011,Teyssier2004}, 
and the Swedish 
ODIN satellite that used a gas cell filled with H$_2$O \citep{Frisk2003}.
Except for the experiment of \citet{Wehres2016}, these experimental setups 
were focussed on the pre-launch characterization of receivers  
and not on the possibility of using { the receivers}
for laboratory spectroscopy and chemical reactors.

The main advantage of thermal emission with respect absorption techniques \citep{Clark1981} is the instantaneous broad band provided by 
radioastronomical receivers 
that permits to follow the variation of intensity with time of the emission lines of many molecular species. The main goal of this proof of concept is to 
show that kinetic simulations (photochemistry, cold plasma, discharges) could be performed addressing chemical compositions similar to those 
of interstellar clouds.
This paper is devoted to the development of a gas cell prototype installed in the beam
path of the receivers of the 40\,m ARIES telescope. It was designed to
study the spectroscopy and chemistry of cold plasmas, with or without the action of UV photons.
Section 2 describes the experimental setup, the data acquisition system, and its performances.
Calibration procedures are explained in Section 3 while in Section 4 we describe
the expected molecular thermal emission in the gas cell. Finally, in Section 5 we present the results obtained
under static or flow modes with, or without, cold plasma or UV irradiation conditions.

\begin{figure}
\label{fg:setup-3}
\centering
\includegraphics[angle=0,width=9.0cm]{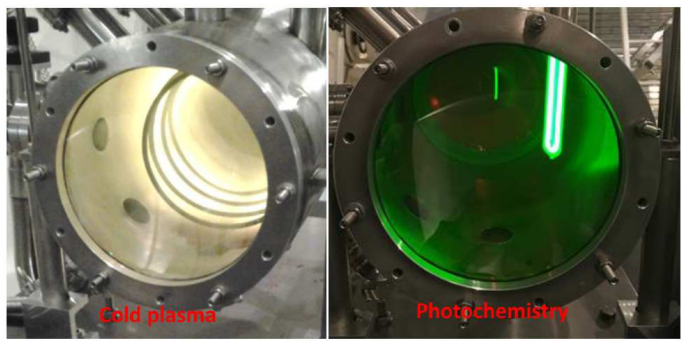}
\caption{
Images of the reactor with plasma on (left) and the UV lamp (right). 
The colours of the plasma glow emission and 
the UV lamp emission are determined by the optical transmission of the
brown Upilex windows.}
\label{Plasma_UV}
\end{figure}

\section{Experimental setup}
\label{sect:setup}

\begin{center}
\begin{figure*}
\label{fg:setup-4}
\includegraphics[angle=0,scale=0.64]{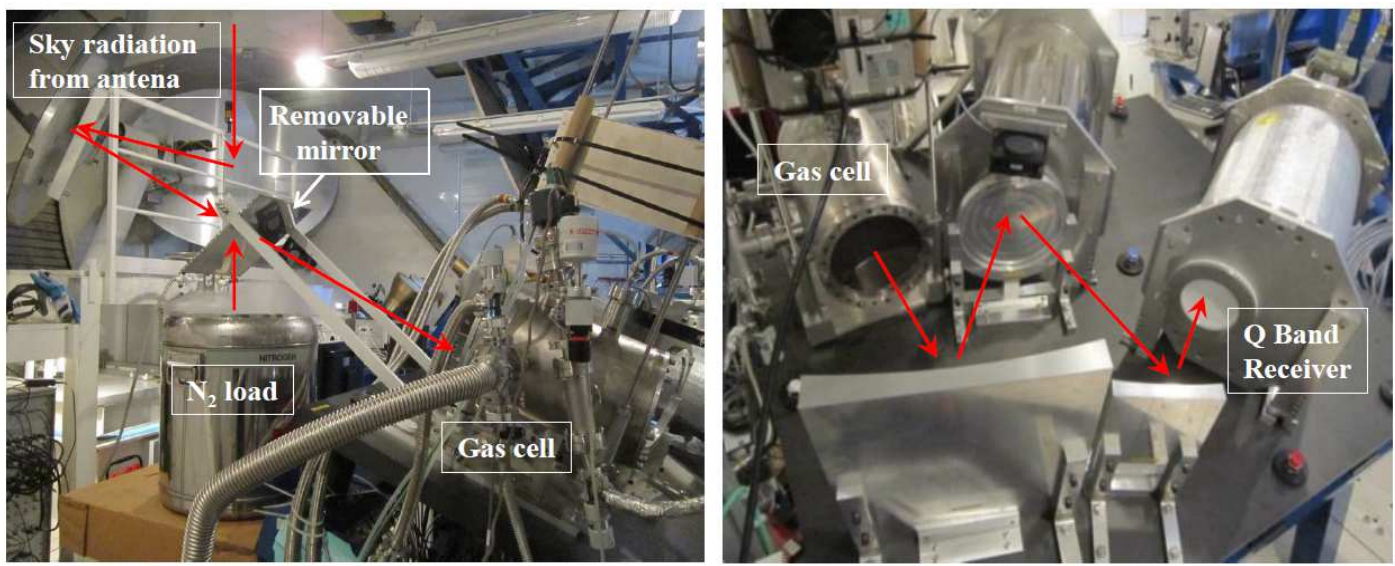}
\caption{Picture of the prototype of the gas cell installed in the receiver room
of the 40\,m IGN radio telescope. The cell is in the optical axis of the Q band receiver.
Most of the
observations were performed using the sky emission as cold load (see text).
In case of poor weather conditions, the N$_2$ cold load was used.
}
\label{installation}
\end{figure*}
\end{center}

\subsection{Basic design}
The basic concept of the gas cell and the detection scheme is depicted in Figure 1. 
{It will be a chamber} of
60\,cm diameter and 100\,cm length 
{that} will be equipped with temperature and pressure sensors, a quadrupole mass spectrometer, pipeline entries for 
the different gases, a cold plasma generator, and UV lamps. 
An absorber at temperature T$_{c}$ (microwave absorber embedded in 
liquid N$_2$ or inside a cryostat at 20 K; hereafter referred to as cold load) will be behind the cell and in 
the direct optical path of the receivers allowing to measure the thermal molecular
emission of the gases inside the chamber (see Section 3). 

A prototype has been built as
described below to demonstrate that radio astronomical receivers can be used for molecular characterization
in experimental gas chambers (spectroscopy and/or chemical reactivity).
The experiments with the gas cell prototype were carried out by installing it in the beam path of the 40\,m 
radiotelescope at CNTRAG-IGN (Guadalajara, Spain). 
Figures~\ref{FigSetup}-\ref{installation}
show the chamber and its installation in the receiver room of the telescope.
The prototype chamber is a 
stainless steel cylinder of 40\,cm length and 25\,cm diameter giving a total volume of 20 liters
approximately. It was designed with non-parallel windows to avoid multiple reflections 
between the two windows during the spectra
acquisition with radio astronomical receivers (see section 2.4). The detection system is mounted on a
table which holds the Q band high electron mobility transistor (HEMT) receiver, the 
mirrors and the cold load. The Q band receiver detects molecular emission from rotational transitions 
in the 41-49\,GHz frequency range against the background radiation coming from the cold load (see Figures 1-4). 
The gas chamber is located where the gaussian beam narrows
to its minimum width to permit the smallest possible size for the windows of the cell.
Regarding the number of ports in the chamber, Figures~\ref{FigSetup}.a
and \ref{FigSetup}.c show the first and the final versions of the gas cell prototype. The 
increase in the number of ports allowed us to improve the flow conditions, and to connect a 
variety of instrumentation. Most of the results presented here were obtained with this final version.  
   
The gas or gas mixture is injected from a gas mixing chamber equipped with six inlets 
with manual dosing valves (four for gases and two for liquids) and a 3 m$^{3}$·h$^{-1}$ rotary pump 
(Pfeiffer Duo 3M) that allows pressure control inside the mixing chamber in the range of
0-100 mbar (Fig.~\ref{FigSetup}.d). Carbonyl sulfide (OCS) from Air Liquide, O$_2$, He,  
and N$_2$ from Praxair
all with > 99\% purity in gas phase, and CS$_2$, CH$_3$OH, CH$_3$CH$_2$OH and HCOOH (Sigma Aldrich > 98\%) in liquid phase, were used in 
this work. In order to avoid liquid condensation
during the injection, the mixing chamber and injection tube were heated up to 40$^{\circ}$C.
   
For static regime experiments the gas is injected from the gas mixing chamber through a manual 
dosing valve until the desired pressure in the range of 0.1-0.5 mbar is reached. However, to carry
out dynamic regime experiments the gas flow and pressure inside the gas chamber were controlled 
at the outlet by a 30 m$^{3}$·h$^{-1}$ dry scroll pump (Agilent Technologies TriScroll 
600). In such conditions, typical gas residence time is approximately 8 s.

The evacuation process between experimental runs is performed with a 250 l/s turbomolecular pump
(Agilent Technologies Twistorr 304FS). 
Initial pumping stages from atmospheric pressure were performed very slowly to prevent the sudden breaking of 
the thin polymeric windows, which deformed inward markedly and in some cases imploded very loudly.
To read the pressure inside the gas chamber in the different
vacuum conditions, two different pressure gauges (capacitance manometer Leybold CTR90
for experiments and Agilent
Technologies FRG-702 Pirani/cold cathode for the initial vacuum) were used, both coupled to a single independent flange 
of the gas-cell through a fine mesh filter and a vacuum tee piece (in the final version of the prototype, see Figure 2c), and 
both connected to their respective gauge 
controllers (Leybold Center Two and Agilent Technologies XGS 600, respectively) in order to register the pressure through an in-house developed software 
routine.

To control the chemical processes during the experiments, a differentially pumped quadrupole mass spectrometer
(Pfeiffer Prisma Plus) working in the 0-100 a.m.u. range is connected to the chamber
through a solenoid actuated regulating gas valve (Pfeiffer RME 005 A valve and RVC-300 controller) that
limits the flow in order to maintain the adequate working pressure for the quadrupole (5·10$^{-6}$ mbar). 
At this working pressure, a Faraday cup was used as detector to avoid saturation or damage of the also 
available Secondary Electron Multiplier (SEM) detector.
We note that the mass spectrometer does not provide information about unstable species, due either 
to their recombination in the path to the spectrometer or because their signals coincide with a 
fragment of a stable species (precursor or product).  Furthermore, it cannot 
discriminate among different species with the same molecular mass.

Depending on the molecular excitation and dissociation process, inductively coupled radio frequency
(RF) discharge or 
UV radiation, different instrumentation was 
placed inside
the chamber. Fig.~\ref{FigSetup}.a and Fig.~\ref{FigSetup}.b (cross section view) show the first 
version of the gas cell with a copper coil
(6\,mm outside diameter copper tubing, four turns, 10\,cm length and 17\,cm diameter) placed inside 
the chamber parallel to the 
primary cylinder axis for low pressure plasma experiments in static or dynamic regime. 
The 
13.56\,MHz RF generator 
(H\"uttinger PFG 300 RF + matchbox PFM 1500A) was connected between one of the ends of the Cu coil (electrically isolated) and the 
grounded chamber at the minimum feasible distance. The shortest possible wires were used to minimize RF radiation noise. 
The other end of the coil was grounded.  RF powers of 5-100 W were typically  applied to sustain the plasma. 
The coil was refrigerated by a circulating water circuit.

For UV radiation experiments, a vacuum system adapted UV lamp (UVB-100, RBD Instruments)
was placed inside the chamber 
where the interference with the receiver beam path is minimized, as shown in Fig.~\ref{FigSetup}.c for the final version of the chamber. The low-pressure Hg lamp spectrum exhibits 
emission lines at 185\,nm (30\%) and 254\,nm (70\%)
with a total power of 20\,W in the UVC range (100-280\,nm).

\subsection{Data acquisition}
\label{sect:dat-ac}
\begin{center}
\begin{figure}
\includegraphics[angle=0,scale=0.31]{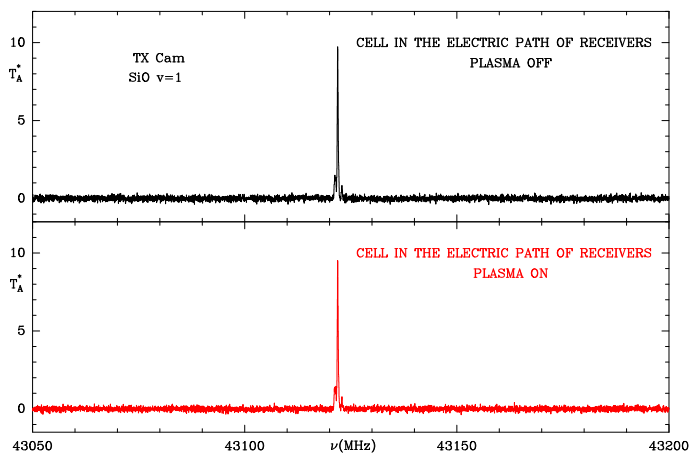}
\caption{Observed spectrum with an He plasma on (bottom) and off (top) towards TX Cam. The selected frequency corresponds
to the SiO $\nu$=1 J=1-0 which is maser in nature in this O-rich evolved star. No spurious signal
potentially produced by the plasma and the associated high-voltage electronics 
were detected. Only 150 MHz of the 2 GHz of instantaneous bandwidth of these observations are shown.}
\label{txcam}
\end{figure}
\end{center}

The prototype of the gas cell was installed in the receiver room of the 40 m IGN radio telescope
as shown in Figure~\ref{installation}. The receiver beam 
width along the axial axis of the cell (at 1/e intensity level) was always smaller than the diameters 
of the gas cell windows and the coil, allowing 
standard radio astronomical observations, if needed, during the experiments
(see below and Figure~\ref{txcam}).
Three different experimental
runs were carried out between October 2015 and November 2016.


Observations were performed with a dual-polarization (left and right circular polarization) 
single-pixel 45 GHz receiver mounted in the Nasmyth focus. The feed horn, orthomode transducer 
and HEMT amplifiers of the receiver were cooled at cryogenic temperature (15 K) in a cryostat. 
The front end provides an instantaneous bandwidth of 9 GHz, from 41 to 50 GHz for each polarization 
but the intermediate frequency (IF) stage limits it to an instantaneous band of 2 GHz. The average 
noise temperature of the receiver is 45 K (the input termination temperature should be added to 
obtain system noise temperature). The design of the IF completely rejects the image frequency and 
the receiver can be considered of the single side band 
(SSB) type. Both base bands from the two 
polarizations, between 0 and 2 GHz, were sent using RF-over-fiber optical links to the backends 
room where the signal was injected into a Fast Fourier Transform Spectrometer with two modules 
of 2.5 GHz band width and 65536 channels, which provides a frequency resolution of 38.1 kHz. 
Only the first 52429 channels (2 GHz) were used in the final spectra.
The Allan variance minimum time, based on the previous astronomical observations,
is of the order of few minutes.

Calibration was performed using a hot and a cold load after having evacuated the gas cell down to a pressure of approximately 
$2\cdot10^{-3}$ mbar. It consisted on two phases, one integration of typically 30 seconds using a 
microwave absorber at room 
temperature at the back of the gas cell and a second integration using the sky as cold load. During 
most experiments the sky was clear and the antenna was pointing at 89 degrees elevation. The estimated cold load temperature was 
obtained for each scan from the weather conditions using 
the atmospheric transmission model (ATM) \citep{Cernicharo1985,Pardo2001} and assuming a forward efficiency 
of 90\% approximately. It was typically 40 K at 43 GHz. We also tested a cold load consisting of a mirror behind the gas 
cell pointing to an absorber inmersed in a liquid nitrogen tank. Results were worse than using the sky since we observed standing waves in the 
reference spectrum. The uncertainty on the calibration has been estimated to be less than 3\%, based on 
the repeatability of the receiver noise temperature after a significant number of measurements across
the receiver bandwidth using the hot and cold loads just in front of the receiver beam.


Most of the observations were performed using the frequency switching technique (FSW) 
with a maximum throw of $\pm$ 24 MHz around the 
central frequency, one second integration per phase and 10 ms for the blank time. Each scan was integrated typically between 
one and three minutes. The calibration applied after doing the substraction between both phases was done at the central 
frequency giving a non flat baseline at the final spectra. In order to get a flatter baseline we used a procedure which 
consisted in doing a frequency switch cycle with the gas cell empty and the cold load, averaging 25 channels around each 
spectral channel and using that averaged reference spectrum to substract it from the calibrated result.

Some observations were done using the standard spatial ON-OFF technique; a typical scan consisted in three phases of 30 
seconds of integration time with the gas cell empty, 60 seconds with the gas cell full and another 30 seconds with the gas 
cell empty. This technique resulted in flatter baselines but it was time consuming and was dependent on gain variations 
with time.  

The frequency switch was used for static or flowing neutral gases while on-off techniques were also 
used when working with the radiofrequency generator that ionized 
the gas in the cell creating a plasma with the off phase corresponding to the plasma switched off. A 
blanking time of a few seconds was introduced between the on and off phases
to allow pumping the by-products of the plasma phase. 

To check that the radiofrequency generator did not inject RFI noise we performed 
several ON-OFF scans, 30 second integration time per phase of a He (0.3 mbar) 60\,W plasma 
being the on phase with the generator 
switched ON and the OFF phase with the generator switched off as during the experiments with other gases. 
System temperature and noise level were not affected by the RF generator.
 Additional tests consisting in observing an astronomical source, TX Cam, which exhibits
a strong maser in the J=1-0 $\nu$=1 line of SiO. The observations were performed before the installation
of the cell, with the cell in the optical path of the receiver but with the plasma generator off, and
finally with the plasma switched on. No spurious signal was detected over the 2 GHz of instantaneous
bandwidth of the spectrometer. Moreover, the intensity of the astronomical line was the same in the
three cases ensuring the total transparency of the gas cell device to electromagnetic radiation
coming from the sky (see Figure~\ref{txcam}).

The setup of the receiver, calibration and observation scans were managed by a software package developed by the authors 
and written in Python, which provides a command line interface. The software monitors the status of all devices, including 
the pressure sensors at the gas cell and the power of the radiofrequency oscillator when used. All parameters were stored 
in a database per scan for logging purposes. Data were recorded in ASCII files and converted on the fly into GILDAS 
compatible spectra and some status parameters were used to populate the headers of each scan. CLASS was later used for 
the analysis (see, http://www.iram.fr/IRAMFR/GILDAS).

\subsection{Spectral sensitivity}
The measurements have been performed 
by frequency switching (for long integrations), or by switching between the cell filled with the experimental gas and 
the cell completely empty (for fast data acquisition),
or by swiching the plasma on and off (see above). 
The switching speed 
depends on the stability of the receivers. The acquistion data rate during the experiment can be as low as 0.1 second 
and depends on the chemical time of the experiments.
For fast chemical experiments the OFF observations will
last a few seconds and will be performed only at the beginning 
and at the end of the experiment.
Hence, the 
noise of the observed spectra is dominated by the ON acquisition time, 
and is given by

\begin{equation}
\sigma=\frac{T_{sys}}{\sqrt{\Delta\nu\,t}}
\label{eq:Sigma_sensi}
,\end{equation}
\noindent
where $T_{sys}$ is the system temperature (in K), $t$ is the integration time (in seconds), and $\Delta\nu$ 
is the spectral resolution (in Hz). 
For a system temperature of 100 K, $t$=1 second, and $\Delta\nu$=0.035 MHz the noise observed in the spectra 
is 0.54\,K.
For optically thick lines the expected intensity is 230 K (see section 3), and for optically thin lines 
it is $\simeq$230$\times\tau$ (see section 3). 
Hence, we expected to measure line opacities $\simeq$ 10$^{-2}$ at 5$\sigma$ in 1 second of 
observing time. We note that for optically thick transitions the system temperature to be
considered for the calculation of the noise in the frequency range covered by the lines
is T$_{rec}$+T$_{cold}$+T$_{gas}$-T$_{cold}$=T$_{rec}$+T$_{gas}$. For the frequencies and gas partial pressure of our 
experiments (see below) all observed lines are optically thin. Hence, the contribution of the lines to the spectral noise in their
frequency ranges is negligible.

The sensitivity of thermal emission spectroscopy is lower than that of CHIRP pulse spectrometers at low
frequency. For frequency observations in the millimeter and submillimeter domain the very large 
instantaneous bandwidth provided by low noise cooled radio astronomical receivers offers a 
sensitivity comparable, and even better, than that provided by standard laboratory absorption 
measurements using warm detectors and narrow frequency bands.

\begin{center}
\begin{figure}
\includegraphics[angle=0,scale=0.34]{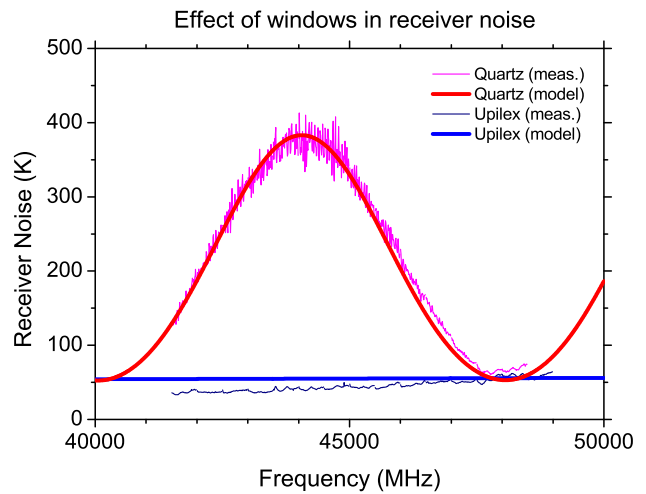}
\caption{Effect on the system temperature of the windows used
in the experiment. The receiver temperature was measured between
41.5 and 49 GHz. The Quartz windows, which are best suited for
high vacuum, have a poor performance and introduce a significant
loss of sensitivity. The Upilex windows have a very good
behaviour concerning transmission but are much more sensitive
to stress when the chamber is evacuated. The thick lines
correspond to the predicted receiver noise temperature by the
model described in the text.
}
\label{windows}
\end{figure}
\end{center}

\subsection{Selecting appropriated windows for the gas cell}
\label{sct:windows-chamber}
Two types of windows were initially considered for the gas cell: a) thick (9.60 mm) fused 
quartz ($\epsilon_r$=3.8) and b) thin (75 $\mu$m) Upilex 
polyimide film ($\epsilon_r$=3.3). The fused quartz windows were initially selected 
for their compatibility with ultra-high vacuum systems, their 
robustness and their chemical inertness, but were expected to reflect a significant part of 
the incoming radiation at some frequencies. On the 
other hand,
Upilex windows with low dielectric constant and small thickness provide an 
almost ideal transmission in the frequency range of interest, but are very fragile and prone 
to degradation by mechanical stress and chemical reactions at their surfaces. 

In order to gain a better understanding of the impact of the electrical properties of the windows on the 
sensitivity of the experiment, a simple model of the combination of the gas cell and the cryogenic receiver was 
elaborated. The windows were simulated as transmission lines with characteristic impedances and 
electrical lengths according to their dielectric constant and thickness. The receiver was assumed to 
present a flat noise temperature of 50 K in the entire band. As the input and output windows are tilted 
by approximately 11$^o$ with respect to the optical axis, it was assumed that no interference exist between them.
However, the interference between the two faces of each window and the reflection of the ambient 
thermal radiation  towards the receiver were included. Figure~\ref{windows} presents the comparison of 
the simple model prediction with the experimental results of the receiver noise temperature obtained by 
measurement with hot-cold load at the input of the gas cell. We note the severe degradation of the sensitivity 
around 44 GHz in the case of the fused quartz windows. 

The effect of dissipative loss in dielectric materials (tan $\delta$) is not obvious from the plot 
shown in 
Figure~\ref{windows} and it is relatively small in the two cases.  Amorphous quartz has very low losses
(tan $\delta\simeq$ 0.0001), but Upilex is slightly worse than other similar polymer materials normally 
used as 
microwave dielectrics (tan $\delta\simeq$ 0.005). To illustrate the effect of dissipative loss in the two materials 
we can compare the estimated added noise due to dielectric loss for the two windows for quartz and Upilex at 
48 GHz (coincident with the minimum reflection of quartz). It is found to be 0.80\,K for Upilex and 1.56\,K 
for quartz. 

Despite the practical problems posed by the thin Upilex windows exposed above, 
most of the experiments were performed with them
due to the their good electrical properties and 
superior sensitivity in the frequency range of interest.
After the completion of
the experiments shown in this paper Teflon windows were {used,} fitted with quarter wave matching layers on both sides in { in order to decrease their reflectivity}. 
The total thickness was 7mm. The matching layers were machined as concentric grooves of 1.4 mm depth and 
1.5 millimeter width spaced by 1.5 mm. The reflection loss of the windows was expected to be below -30 dB 
in the whole band of the receiver. The noise contribution due to dielectric dissipative loss has been measured
and found to be negligible. Teflon offers superior safety conditions than Upilex 
and similar transmission performances.

\section{Absolute data calibration}
Let us assume that the windows of the cell are transparent at microwave wavelengths. The receivers are equipped 
with HEMTs (see below) so that the data are in single side band mode and we do not have to deal with
the always uncertain gains of upper and lower sidebands typical in SIS superheterodyne receivers. 
Facing the receiver, behind the cell, there is a black body at temperature $T_{c}$ (cold load, see Figure 1). 
The gas in the cell is at temperature $T_g$ with a volume density high enough to maintain it under thermal
equilibrium conditions, that is,
all rotational levels are populated following a Boltzmann distribution at a single
rotational temperature which is identical to the kinetic temperature of the gas.
As frequencies in our experiment are below 50 GHz and the physical temperature of the cell is $\sim$300 K, it turns out
that $h\nu/k_BT<\,$0.01 ($h$ and $k_B$ are the Planck and Boltzmann constants respectively). In those conditions, we can
assume the Rayleigh-Jeans approximation {\textbf{\color{blue}applies}} to the line flux $I_\nu$ and use a brightness temperature, $T_B$,
so that $I_\nu(T)=2\,\nu^2\,k_B\,T_B/c^2$.  Hence, the line intensity can be given as a temperature so that the signal
detected by a receiver in front of the cell at a frequency $\nu$ (assumed to be close to the central
frequency $\nu_{ul}$ of the molecular transition $u \rightarrow l$) can be written as

\begin{equation}
T_B^{ON}(\nu)=T_c\,e^{-\tau(\nu,T_g)}+T_g\,(1-e^{-\tau(\nu,T_g)})=(T_c-T_g) e^{-\tau(\nu,T_g)}+T_g
\label{eq:TB_ON_1}
,\end{equation}

\noindent
where $\tau(\nu,T_g)$ is the gas opacity integrated along the cell length at frequency $\nu$ and temperature
$T_g$ (see below). When $T_{c}$=$T_g$, there is a lack of contrast and therefore the signal detected will not display
any spectral signature: $T_B(\nu)$=$T_g$. 
In this consideration we have neglected the emissivity and
absorption of the cell windows (see below). 
The opacity of the line depends on the dipole moment of the molecule, the total number
of molecules along the path, the frequency, and the temperature (through the partition function) 
and will be discussed in the next section. When the receiver observes through an empty cell 
the detected signal will be 

\begin{equation}
T_B^{OFF}(\nu)=T_c
\label{eq:TB_OFF_1}
.\end{equation}

The signal $T_B^{ON-OFF}(\nu)$, obtained by the standard ON-OFF technique in radioastronomy will be given by

\begin{equation}
T_B^{ON-OFF}(\nu)=(T_g-T_c)\,(1-e^{-\tau(\nu,T_g)})
\label{eq:TB_ON-OFF_1}
.\end{equation}

If $T_g>T_c$ we will observe the spectral lines in emission. If $T_g<T_c$ they will be in absorption. For practical 
purposes we can assume $T_c$=70 K and $T_g$=300 K. 
In the optically thick case, that is, $\tau$$>>$1, $T_B^{ON-OFF}$=230 K. In the optically thin case,
$\tau$$<<$1, $T_B^{ON-OFF}$=230\,$\tau(\nu,T_g)$\,K.
In the above equations we have assumed that air is practically transparent at the wavelengths
of the experiment. However, O$_2$ has a cluster of absorption lines in the range 50-70 GHz and another one centred
at 118.75 GHz. Water 
has an absorption feature at 22 GHz. Hence, the air path in this experiment should be maintained as short as possible.
Nevertheless, the absorption produced by the air can be corrected using the ATM model \citep{Cernicharo1985, Pardo2001}.  

A more tricky problem is the absorption by the material used for the windows of the cell. As, in general,
that absorption will not be negligible, we can generalize Equation \ref{eq:TB_ON_1} as follows:
\\
\\
$T_B^{ON}(\nu)=[\{T_c\,e^{-\kappa_\nu}+T_w\,(1-e^{-\kappa_\nu})\}\,e^{-\tau(\nu,T_g)}+$
\begin{equation}
\;\;\;\;\;\;\;\;\;\;\;\;\;\;\;+T_g\,(1-e^{-\tau(\nu,T_g)})]\,e^{-\kappa_\nu}+T_w\,(1-e^{-\kappa_\nu})
\label{eq:TB_ON_2}
,\end{equation}
\\
\\
\noindent
where $T_w$ is the temperature of the windows (in principle identical to $T_g$), and $\kappa_\nu$ is the 
opacity of the windows at frequency $\nu$. 
The first term represents the emission of the cold load attenuated by the first window. The second term 
represents the emission of the first window. 
Both terms are attenuated by the gas in the cell. The third term represents the emission of the gas. 
Finally the window in front of the receiver will attenuate all the 
previous terms and emit itself (last term). In case of an empty cell, we get the OFF signal:

\begin{equation}
T_B^{OFF}(\nu)=[T_c\,e^{-\kappa_\nu}+T_w\,(1-e^{-\kappa_\nu})]\,e^{-\kappa_\nu}+T_w\,(1-e^{-\kappa_\nu})
\label{eq:TB_OFF_2}
,\end{equation}

\noindent
and the detected ON-OFF signal will be

\begin{equation}
T_B^{ON-OFF}(\nu)=[T_g-T_c\,e^{-\kappa_\nu}-T_w(1-e^{-\kappa_\nu})](1-e^{-\tau(\nu,T_g)})\,e^{-\kappa_\nu}
\label{eq:TB_ON-OFF_2}
.\end{equation}

The absolute calibration is obtained, as mentioned in section \ref{sect:dat-ac}, through the
use of two microwave absorbers at different temperatures (liquid N$_2$, or clear sky, and ambient).
The detected flux, $W$, is transformed into temperature, $T$, through

\begin{equation}
T=W\,(T_h-T_c)/(W_h-W_c)
\label{eq:HOT-COLD}
,\end{equation}
  
\noindent
where $W_h$ and $W_c$ are the signals detected by the receiver from the hot and cold loads, $W$ is the 
flux detected in a given observation phase (ON, OFF, or FSW [frequency 
switching]), and $T_h$ and $T_c$ are the temperatures of the hot and cold loads.

The accuracy of the calibration can be degraded by the atmosphere, the uncertainty in the
transmission of the cell windows, as well as instrumental baseline effects. The short atmospheric
path involved in the experiment ensures that the 
atmosphere is not perturbing the observations. 
Taking this factor into account, we can estimate uncertainties below 5\% in our measurements
(to be compared with the 3\% from the repeatability obtained in the measurement of the receiver
noise temperature). Hence, the partial pressures, or molecular abundances, of the gases in the cell 
derived from the observed intensities of their rotational lines will exhibit similar error bars
($\simeq\pm$5\%).

As the observation frequency increases, the Rayleigh-Jeans approximation starts to fail.
Consequently, in the submillimeter and far-infrared domains
$I_{\nu}$ has to be replaced in the calibration equations by the full Planck function, $B_\nu(T)$, 
given by

\begin{equation}
B_\nu(T)=(2\,h\,\nu^3/c^2)\,(e^{(h\,\nu/k_B\,T)}-1)^{-1}
\label{eq:planck}
,\end{equation} 

\noindent
where $\nu$ is the frequency, h is the Planck constant, c is the speed of light, and 
$k_B$ is the Boltzmann constant.

\begin{center}
\begin{figure}
\includegraphics[angle=0,scale=0.32]{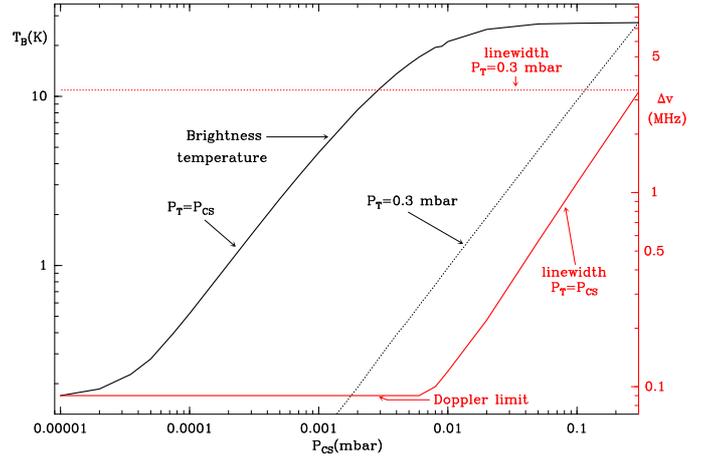}
\caption{Expected brightness temperature (black lines) and linewidth (red
lines) for the CS J=1-0 line
for two different cases. The continuous lines correspond to the cell filled
only with CS and P$_T$=P$_{CS}$, while the dashed lines correspond to a cell
filled with N$_2$ and CS at a total constant pressure of 0.3 mbar. The x-axis
corresponds to the partial pressure of CS.}
\label{cs_teo}
\end{figure}
\end{center}

\section{Molecular thermal emission in the gas cell}
The volume density, $n$, of a given gas in molecules cm$^{-3}$ (hereafter these units are referred to 
as cm$^{-3}$) in the cell is given by

\begin{equation}
n=7.2431\cdot10^{18}\,P\,/T
\label{eq:dens}
,\end{equation}

\noindent
where $P$ is the partial pressure (in mbar) and $T$ the kinetic temperature (in K) of the gas. 
The gas column density, $N$, in cm$^{-2}$ along the optical path $d$ (in cm) of the experiment is given by

\begin{equation}
N=7.2431\cdot10^{18}\,d\,P\,/T
\label{eq:col-dens}
,\end{equation}


The opacity of a molecular line corresponding to a transition $u \rightarrow l$  is given by

\begin{equation}
\tau(\nu,T_g)=c^2\,N\,g_u/(8\pi\,Z\,\nu^2_{ul})\,\times\,e^{-E_l/T_g}\,A_{ul}\,(1-e^{-h\nu_{ul}/k_BT_g})\,\phi(\nu)
\label{eq:op-line}
,\end{equation}
\noindent
where $Z$ is the partition function, $A_{ul}$ is the Einstein coefficient of the transition, $g_u$ is the
degeneracy of the upper level, $E_l$ the energy of the lower level of the transition (in K), and  
$\phi(\nu)$ is the normalized line profile. 
The Einstein coefficient is given by

\begin{equation}
A_{ul}=\frac{64\,\pi^4}{3\,g_u\,h\,c^3}\,\,\nu_{ul}^3\,\mu^2_{ul}
\label{eq:eins-coef}
,\end{equation}

\noindent
where $\mu^2_{ul}$ is the square of the transition dipole moment which, in turn, can be 
written as S$_{ul}$\,$\mu^2$, where S$_{ul}$ is the so-called transition line strength and $\mu$ is
the permanent dipole moment of the molecule. Therefore, the line opacity can be written as follows:
\\
\\
$\tau(\nu,T_g)=(8\,\nu_{ul}\,\pi^3/3\,h\,c)\,(NS_{ul}\,\mu^2/Z)\,(e^{-E_l/T}-e^{-E_u/T})\,\phi(\nu)=$
\begin{equation}
\;\;\;\;\;\;\;\;\;\;\;\;\;\;\;\;\;\;=I_{ul}(T)\,N\,\phi(\nu)
\label{eq:line-op}
.\end{equation}

\noindent
The parameter I$_{ul}$ is provided for a significant number of molecules in the JPL 
\citep{Pickett1998} and CDMS \citep{Muller2005} line catalogues. The MADEX code
\citep{Cernicharo2012} also provides this value for more than 5600 molecules
when used in its LABO mode.

The thermal or Doppler ($\Delta\nu_D$) line width (half width at half maximum, HWHM), 
for a molecule of mass $M$ (in atomic units) is given by\begin{equation}
\Delta\nu_D=3.581\cdot10^{-7}\,\nu_{ul}\,(T/M)^{1/2}
\label{eq:doppler-width}
.\end{equation}

The Doppler line profile can be assumed to be a Gaussian with a HWHM of $\Delta\nu_D$.
The spectral signal will be maximum at the line centre (corresponding to the maximum absorption 
coefficient per unit length, $\alpha_{max}$; $\tau_{max}=\alpha_{max}\,\times\,d$).  
In the so-called Doppler limit it is given by 

\begin{equation}
\tau_{max}(T)=I_{ul}(T)\frac{T_0}{T}\frac{P}{\Delta\nu_D}\times151.194 \times d
\label{eq:doppler-limit}
,\end{equation}
\noindent
where $I_{ul}(T)$ is in units of nm$^2$\,MHz,
$P$ is the partial pressure of the gas in Torr, $\Delta\nu_D$ is the
Doppler HWHM in MHz, and $d$ the length of the cell in cm, and $T_0$=300 K.

Pressure or collisional broadening will induce line profile with a HWHM given by  
$\Delta\nu_C$=$\Delta\nu_C^0$\,$P$, where $P$ is the pressure and $\Delta\nu_C^0$ is the broadening
induced at a reference pressure (normally 1 Torr; 1 Torr = 1.33322 mbar).

Above a certain pressure the line profile will be dominated by the effect of collisions on the width of the
energy levels involved in the transition. The profile $\phi(\nu)$ of a 
collisionally broadened line can be approximated by the Van Vleck-Weisskopf (VVK) 
profile (see Appendix A). When the line is collisionally broadened the absorption
coefficient $\tau_{max}$  does not depend on the pressure as both, the broadening and
the number of particles have the same dependence with the pressure. 
The value of $\tau_{max}(T)$ at temperature $T$ is given by:

\begin{equation}
\tau_{max}(T)=I_{ul}(T)\frac{T_0}{T\,\Delta\nu_C^0}\times102.458 \times d
\label{eq:picket}
,\end{equation}

In this expression $\Delta\nu_C^0$ is in MHz/Torr.
If $\Delta\nu_C^0$ is given in MHz/mbar then

\begin{equation}
\tau_{max}(T)=I_{ul}(T)\frac{T_0}{T\,\Delta\nu_C^0}\times76.850 \times d
\label{eq:picket2}
,\end{equation}

\noindent 
(Equations \ref{eq:doppler-width}-\ref{eq:picket2} are taken or adapted from \citet{Pickett1998}). To illustrate this, we show in Figure~\ref{cs_teo} the expected brightness temperature of the
CS J=1-0 line for a cell length of 40\,cm, ambient temperature of 300\,K, and cold load temperature of 70\,K.
Two cases have been considered, the cell filled only with CS ($P_T=P_{CS}$)
and the cell filled with CS and another gas (N$_2$) with a total pressure, $P_{CS}$+$P_{N_2}$, of 0.3 mbar. Brightness temperatures for CS $J=1-0$ can be obtained using 
I$_{1-0}(300{\rm K})=2.381\cdot10^{-4}$ nm$^2$\,MHz \citep{Muller2005}, 
and Equation \ref{eq:TB_ON-OFF_2}. 
In the case $P_T=P_{CS}$=0.3 mbar, the volume density of CS is 7.22\,10$^{15}$ cm$^{-3}$,
and the corresponding column density in the cell is 2.89\,10$^{17}$ cm $^{-2}$.
The HWHM of CS has been assumed to be 7.5 MHz at 1 Torr for CS-CS self broadening and CS-N$_2$
broadening. Hence, for the case of constant total pressure of 0.3 mbar the linewidth of the observed
transition should be 3.38 MHz.
The maximum brightness temperature that we could expect in the case $P_T=P_{CS}$=0.3\,mbar for this CS line is 27.3 K and will
be obtained for a CS partial pressure of $>$0.05\,mbar. In the case that the partial pressure of CS is low compared with
the total pressure (dashed lines) the width of the transition will be always nearly the same, that is, 3.38 MHz. However,
the intensity varies linearly with pressure from 27.3 K for $P_{CS}$=0.3\,mbar and $P_{N_2}$=0, to 0.2 K for
$P_{CS}\simeq$0.007\,mbar and $P_{N_2}$=0.293\,mbar.

It must be noted that $\Delta\nu^0_C$ also depends on the temperature, 
through a relation usually expressed as 
\begin{equation}
\Delta\nu_C^0(T)=\Delta\nu_C^0(T_0)\times\left(\frac{T_0}{T}\right)^n
,\end{equation}
where $T_0$ is a reference temperature, $n$ is determined empirically for each 
molecule, being typically  $n\simeq0.7-1$. 
See Appendix A for more details on the line profile and the effect of broadening on the
intensity.

\begin{center}
\begin{figure}
\includegraphics[angle=0,scale=0.345]{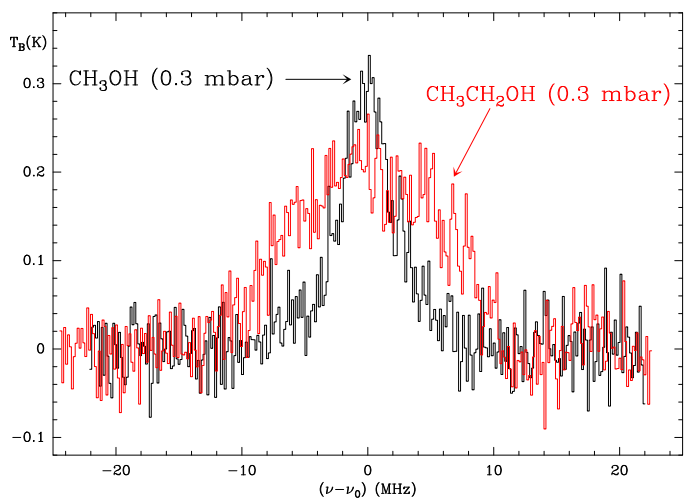}
\caption{Observed lines of methanol and ethanol at 44069 and 46980 MHz
respectively.}
\label{c2h5oh_ch3oh}
\end{figure}
\end{center}

\section{Results}

\subsection{Static or flowing gas}
In the first experiments we selected ethanol and methanol at different pressures as test molecules,
because of their many lines in the Q band.  The temperature inside the gas cell was the ambient temperature
in the receiver room of the telescope and close to 300 K.
However, 
these molecules have large partition functions and self-broadening 
coefficients at 300 K, both factors playing against the peak intensity of 
the spectral lines we could detect.   
Methanol has two symmetry species, A \& E, and ethanol two conformers. Moreover,
at 300 K several vibrational and internal rotational modes are
populated, also decreasing the expected intensities of the rotational lines
at their ground states. For example, for CH$_3$OH, the expected peak intensity of its
line at 44069 MHz 7$_{0,7}$-6$_{1,6}\, A^+$ transition, I(300\rm K)=8.801$\cdot10^{-7}$ nm$^2$\,MHz) 
calculated by MADEX \citep{Cernicharo2012}
is only 0.3 K. For the 46980 MHz ethanol line (7$_{1,6}$-7$_{0,7}$ transition of the 
{\it trans} conformer, 
I(300\,K)=3.352$\cdot10^{-6}$ nm$^2$\,MHz), the expected
peak intensity is 0.2 K for the conditions of the experiment.
The first experimental run was held in October 2015 using the quartz windows. 
To obtain the spectra we performed
ON/OFF scans, according to Equations \ref{eq:TB_ON-OFF_1} and \ref{eq:TB_ON-OFF_2},
with ON: chamber filled with the gas, and OFF: empty chamber. However,
no lines were observed.

\begin{center}
\begin{figure}
\includegraphics[angle=0,scale=0.65]{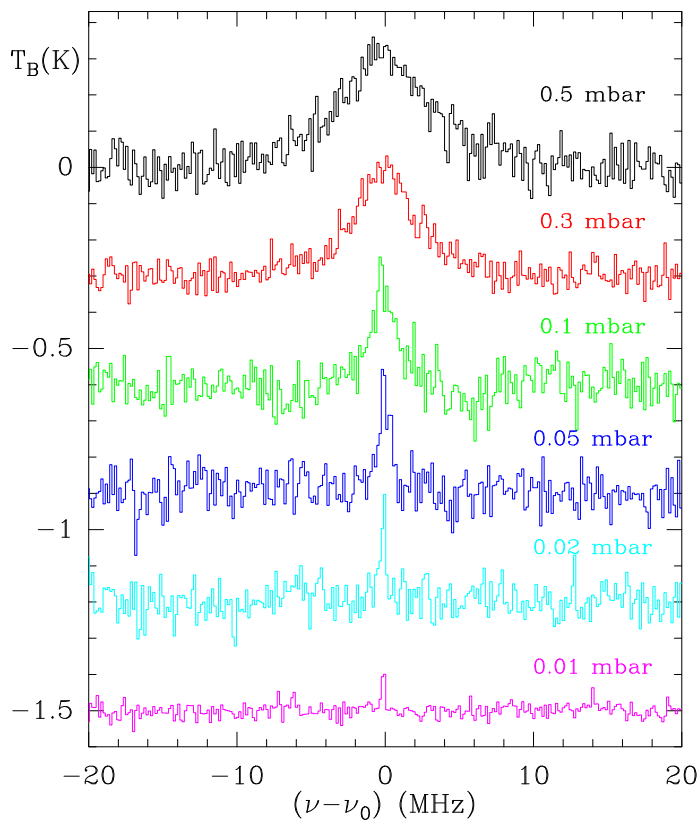}
\caption{Observed line of methanol at 44069\,MHz 
(7$_{0,7}$-6$_{1,6}\, A^+$ transition)
for different
values of the total gas pressure. The integration time was 40 min for the narrower feature. The
spectral resolution is 180 kHz obtained by a five channel smoothing of the original  
data. A shift by -0.3 K is introduced in the vertical scale of each spectrum with respect
to the previous one, as pressure decreases.}
\label{ch3oh_pressures}
\end{figure}
\end{center}

\begin{center}
\begin{figure}
\includegraphics[angle=0,scale=0.49]{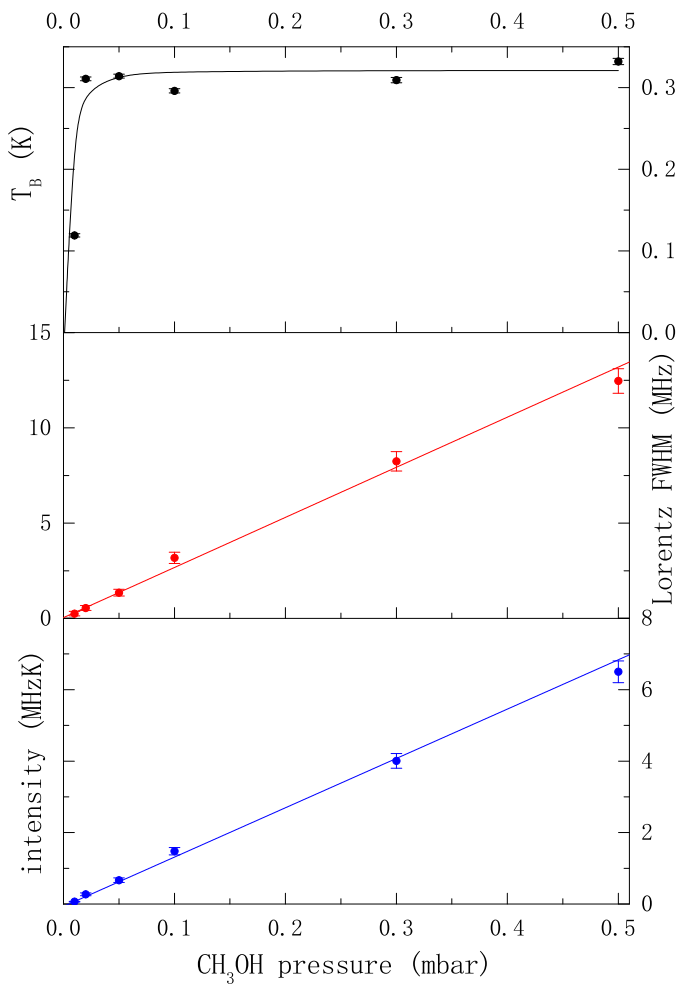}
\caption{
Symbols (from top to bottom): peak intensity, collisional linewidth and
integrated intensity derived from Voigt fits to the observed methanol line
(7$_{0,7}$-6$_{1,6}$, A$^+$); data from Figure~\ref{ch3oh_pressures}).
In all fits the Gaussian width was fixed to the calculated value. Solid lines
(from top to bottom): calculated peak intensity with the calculated broadening coefficient,
and intensity, best fit of the collisional widths vs. pressure (the slope yields a broadening coefficient
$\Delta\nu_C$=13.2$\pm$0.6 MHz/mbar) and best fit of the
integrated line intensity vs pressure.
}
\label{ch3oh_plot}
\end{figure}
\end{center}

The problem arose from the quartz windows which reduce the 
sensitivity (see Figure~\ref{windows}) and, in addition, introduce baseline artefacts due to internal reflections.
We have also noticed that thermal emission from people around the chamber
leads to significant degradation of the baseline. 
In the following experiments we installed the Upilex windows which have a much
better transmission (see section \ref{sct:windows-chamber}).
Figure~\ref{c2h5oh_ch3oh}
shows the lines of CH$_3$OH and CH$_3$CH$_2$OH after 10 min of ON-OFF observing time
at a total static pressure of 0.3 mbar when the chamber was filled with each of these gases.
From just a single measurement of the ethanol line we could estimate a broadening
coefficient of 20 MHz/mbar.
Figure~\ref{ch3oh_pressures} shows the methanol line observed at different
pressures (baseline ripples can be seen in the 0.1 mbar spectrum). As pointed out in
the previous section, in the pressure broadened regime, the peak height remains constant
while the linewidth changes with pressure. For pressures below 0.02 mbar the line
reaches its Doppler width (see Figure~\ref{ch3oh_plot}).
Typical integration times were 10 min during
these measurements except for the lowest pressure for which we integrated during 40 min. 
All spectra were fitted to Voigt profiles, with the Doppler width fixed to the expected
value at 300 K (97 kHz), and peak intensities, collisional linewidths (FWHM) and
integrated intensities where derived from the fit. Figure~\ref{ch3oh_plot}
shows these experimental
values versus pressure, and the linear fits that show the linear dependence of
collisional width and integrated intensity with pressure. The self-broadening
coefficient thus derived is 13.2$\pm$0.6 MHz/mbar (HWHM). The upper panel of Figure~\ref{ch3oh_plot} 
also shows the expected evolution of the peak intensity using the derived
broadening parameter.

The Upilex windows were adopted for all the other observing periods. 
In the second observing run in April 2016 we used
HCOOH and OCS
as probe molecules for static or flowing conditions. 
The observed baselines were very good for standard ON/OFF observations but
exhibited still strong baseline ripples in FSW mode. However,
we noticed that the rippled instrumental baseline was constant in time when
the sky was used as cold load.
Hence, an instrumental baseline was derived before filled-cell scans by
integrating with the empty cell for 40 seconds in FSW mode. This curve was then
subtracted to all subsequent observations resulting in very flat baselines.

\begin{center}
\begin{figure}
\includegraphics[angle=0,scale=0.355]{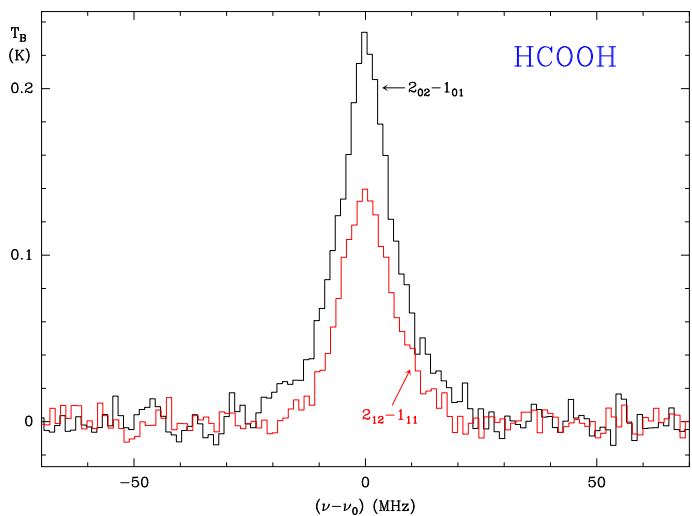}
\caption{Observed emission of the $2_{12}-1_{01}$ and $2_{02}-1_{01}$ trans HCOOH transitions  
at 43303 and 44911 MHz respectively. The integration time was 4 min. Both lines were observed simultaneously.
The spectral resolution is 540 kHz after a channel smoothing of the original data.}
\label{hcooh}
\end{figure}
\end{center}

The partition function of OCS is smaller than that of methanol and ethanol at 300 K.
For HCOOH (formic acid) the partition
function is slightly smaller than that of methanol and the expected intensities in
the 41-49 GHz range are of the same order than those of methanol and ethanol. Figure~\ref{hcooh}
shows two lines of the trans conformer of formic acid observed simultaneously.  
The observing time in these observations was 4 min.

\begin{center}
\begin{figure}
\includegraphics[angle=0,scale=0.355]{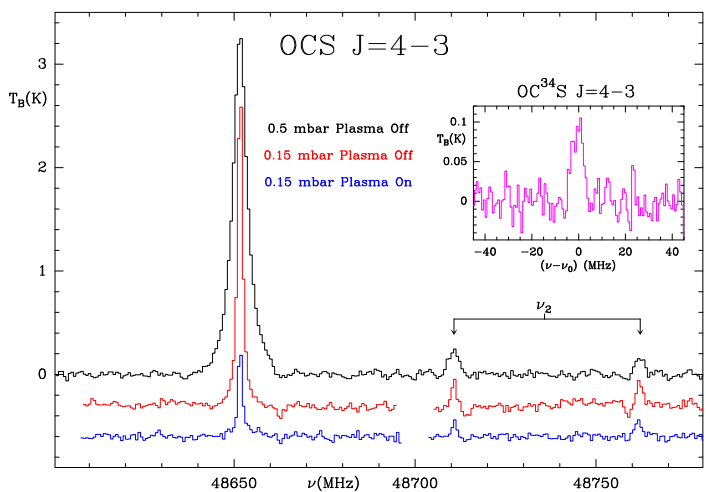}
\caption{Observed emission of the J=4-3 OCS rotational line for different experimental conditions.
The doublet of the $\nu_2$ J=4-3 transition is also detected in
the experiments. The top spectrum was taken in ON/OFF mode, that is, signal through OCS filled cell 
minus signal through empty gas cell. The other spectra were acquired in FSW mode. The
blanked channels in the baseline correspond to the negative features produced by
the frequency folding of the spectra. The insert shows the J=4-3 line of
OC$^{34}$S at 47462.353\, MHz for $P$(OCS)=0.5\, mbar.}
\label{ocs_v2}
\end{figure}
\end{center}

Figure~\ref{ocs_v2} shows the observed line profile of OCS J=4-3 for different experimental
conditions. The bottom spectrum corresponds to the plasma experiment and will be discussed
below. Total observing time is around 10 min and the spectra have been smoothed to a spectral
resolution of 1.0 MHz. The spectra show the rotational emission from OCS in the ground state
and in its $\nu_2$ mode, placed at 749\,K above the ground state. The observed intensities
are those expected for a gas at 300\,K. During the experiment with a cell pressure of 0.5 mbar
it was possible to detect also the J=4-3 transition of the OC$^{34}$S isotopologue (see insert
in Figure~\ref{ocs_v2}). 
From the observed spectra in Figure~\ref{ocs_v2} we derive
$\Delta\nu_C$ (HWHM), of 6.0$\pm$0.5 MHz/mbar,
in good agreement with \citet{kos09}.

When the gas cell is filled with OCS at a pressure of 0.3 mbar, and adopting 
the pressure broadening coefficient derived above, $T_c$=70\,K, and $T_g$=300\,K, then
the MADEX code predicts
an intensity for the OCS $J$=4-3 transition at 48651 MHz of $\simeq$3.3\,K (see Appendix A).
The observed OCS intensities, as shown in Figure~\ref{ocs_v2}, are in agreement with these predictions.
Hence, the partial pressure of the gases in the
cell can be derived with high accuracy from the measurement of the thermally populated 
rotational lines. As expected for these volume densities no differences were found between the 
static and flowing gas case for identical pressures of the gas in the cell.

\begin{center}
\begin{figure}
\includegraphics[angle=0,scale=0.35]{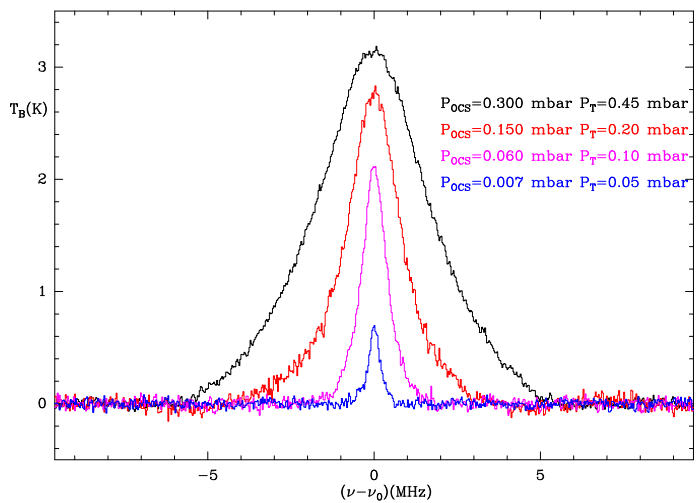}
\caption{Observed J=4-3 OCS transition at 48651.604\,MHz for different values of OCS and total pressures.
The integration time was 4 min for the lowest pressure. The spectral resolution is 30\,kHz.}
\label{ocs_pressures}
\end{figure}
\end{center}

Observations of the $J$=4-3 line of OCS at different partial and total
pressures (in static conditions, with varying contributions of background air) are shown in Figure~\ref{ocs_pressures}.
The mixing of OCS and air during this experiment was not well under control. 
Hence, we estimate that the total and partial pressures shown in
Figure~\ref{ocs_pressures} have uncertainties of $\simeq$30\%.

\begin{center}
\begin{figure}
\includegraphics[angle=0,scale=0.35]{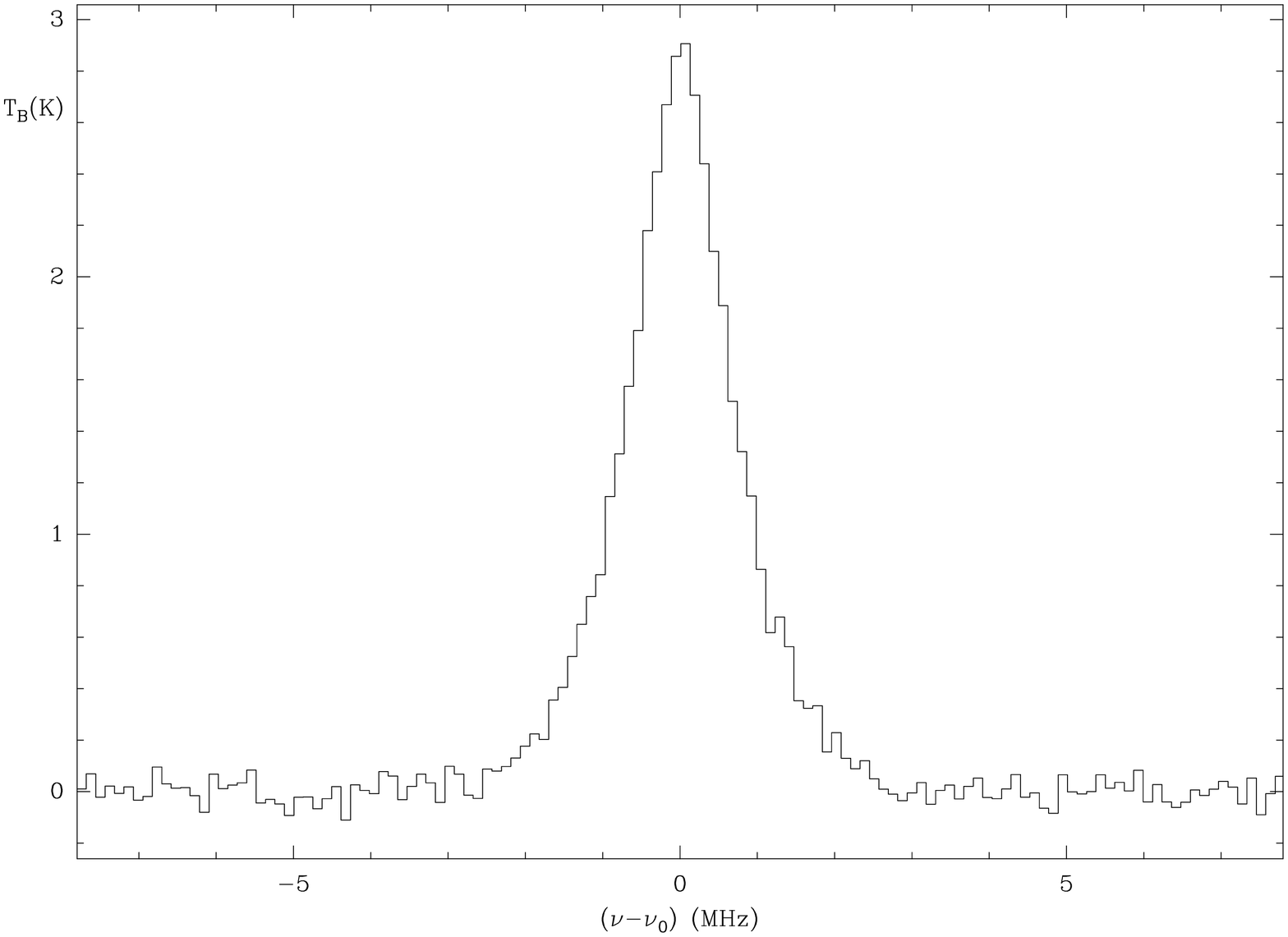}
\caption{Observed CS $J$=1-0 signal (centred at 48990.957\,MHz) at 0.15 mbar and 50 W OCS plasma.
The integration time was 2 min. The spectral resolution is 160\,kHz.}
\label{cs_plasma}
\end{figure}
\end{center}

\begin{center}
\begin{figure}
\includegraphics[angle=0,scale=0.35]{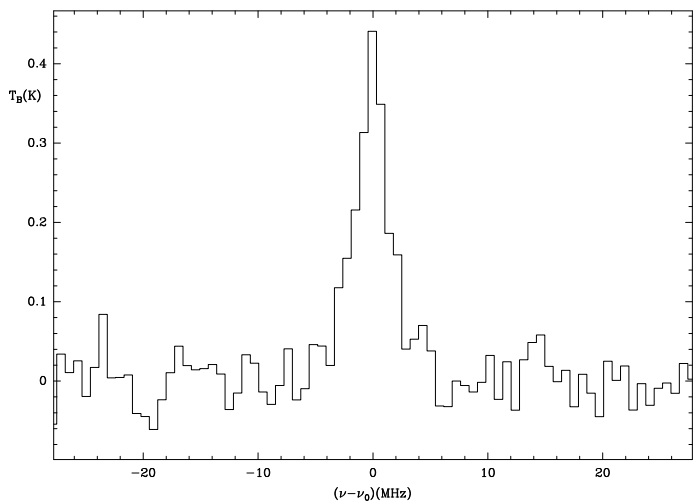}
\caption{Observed SO$_2$ 14$_{2,12}$-13$_{3,11}$ signal (centred at 47913.427\,MHz) 
during the cleaning of the cell through an O$_2$ plasma at 0.1 mbar.
The integration time was 5 min. The derived partial pressure of SO$_2$ is 0.04 mbar (see text).The
spectral resolution is 640\,kHz.}
\label{so2_cleaning}
\end{figure}
\end{center}

\subsection{Cold plasma conditions}
The prototype of the gas cell was equipped with the cold plasma generation described in
Section \ref{sect:setup} (see Figures 2-4). In this type of
glow inductively coupled plasma (ICP) discharges the typical electron 
densities are 10$^9$-10$^{10}$ cm$^{-3}$, while typical electron
temperatures are 3-5 eV (assuming Maxwellian distributions). The most energetic electrons, 
with energies spanning up to several
tens of eV, are responsible for the primary electron impact dissociation and 
ionization processes that begin the rich chemistry at low temperatures
associated to this kind of plasmas.

First studies on the kinetics of cold OCS and CS$_2$ plasmas produced at low pressure 
were carried out in 1980s by different authors.  
\citet{Clark1981} employed millimeter and submillimeter rotational spectroscopy, based on klystron technology, to show that
OCS decomposes very efficiently in DC plasmas discharges through the cleavage of the C=S double bond, resulting in the formation
of CO + S, while more complex reaction pathways resulted in small concentrations of CS and SO, and the excess sulphur was deposited on the reactor
surfaces. ICP discharges of OCS were studied by \citet{Bezuk1983} using mass spectrometry and
optical emission spectroscopy. 
CO was found to be the major product, but S and S$_2$ were detected too. 
According to these authors, CS is formed in its $^1\Pi$ excited state through
electron collisions with OCS and OCS$^+$.

%

\citet{Nicholas1982}  studied OCS and
CS$_2$ decomposition in a RF discharge. Their results corroborated the primary dissociation steps of OCS reported by previous authors, and showed
that the main dissociation pathway for CS$_2$ should be e$^-$ + CS$_2$ $\rightarrow$ CS + S + e$^-$. More recent studies on CS$_2$ plasmas have been
devoted to their application as sources of functional polymers \citep[and references therein]{Ochiai2005} or as a method for CS$_2$ waste removal
\citep{Holub2014,Yan2013, Tsai2007}.

We produced cold plasmas of pure  OCS and pure CS$_2$ at 0.15 mbar. In the case of OCS, different RF powers
(20-100 W) in gas flow conditions produced steady state concentrations of the precursor between 40\% and 14\% of
its initial value. The steady state concentration of CS$_2$ with a 50\,W discharge was 40\%. Mass spectra 
showed the formation of CO in OCS plasmas, as the major stable by-product. S was also
observed as a minor product.
In both plasmas, the CS radical was detected through its rotational emission in the gas phase
with a S/N$>$10 in a few seconds of integration time (see Figure~\ref{cs_plasma}).
The observation of CO and CS in these plasmas is consistent
with previous studies discussed above.
For static gas plasma, the precursor and the products were quickly destroyed, a fact 
confirmed by the pressure decrease in the chamber and by the mass spectrum. We could,
however, observe CS during several seconds before the precursor was completely destroyed. 
In this case the pressure decrease is related to deposition of the gas phase particles on the 
walls. In fact, a very thin
layer of material was observed after a few minutes on the walls and the windows. 
Moreover, dust particles were observed easily
floating in the plasma through the scattering of a laser beam inside the chamber. 
In flowing gas conditions the walls were also covered by a thin layer of sulphur compounds but at lower
rate than in the static case. In this case the CS line was observed maintaining its intensity
during the experiment.

Figure~\ref{cs_plasma} shows the CS line observed in a flowing OCS plasma for a long integration
of 120
seconds. Similar results were obtained with the plasma of CS$_2$. Figure~\ref{ocs_v2} shows
the $J$=4-3 line of OCS with the cell filled with OCS at 0.15 mbar before (red spectrum) and after
(blue spectrum) plasma ignition. The line intensity decreases as well as the linewidth because of
the drop in the total pressure of the chamber. Moreover, the two lines of the $\nu_2$ vibrational
mode decrease less than the line from the ground state, from which we derive that the kinetic temperature
in the chamber increases from 300 to $\simeq$450\,K. 
This demonstrates that the HEMT receivers are sensitive enough
to trace low abundance species and to monitor the kinetic temperature of the gas through the
observation of the vibrationally excited modes of the molecules in the cell.
For example, with a total observing time of 10 min, lines as weak as 0.1 K could be detected with a 
signal to noise ratio of seven as it is the case of OC$^{34}$S shown in Figure~\ref{ocs_v2}. The 
observation corresponds to a partial pressure of OC$^{34}$S of 0.025 mbar and a total 
pressure of 0.5  mbar. The detection limit will depend on the dipole moment of the molecule,
the partial and total pressures, and the line strength of the observed transition.

It is worth mentioning the great difficulty to monitor the CS reactive species by differentially pumped
mass spectrometry, due to its loss of concentration by recombination in the path to the mass spectrometer and 
since most of the detected signal at mass 44 belongs to the fragment produced by electron impact on CS2 or OCS in the ionization 
chamber of the mass spectrometer. Thus the information provided by the emission spectrum is highly valuable.

We also note that, when only one emission line is observed for a given species, there
might some uncertainty in the partial pressure derived from the emission measurements,
due to the possible different rotational and vibrational temperatures.  Although 
rotational temperatures are most likely equilibrated at the kinetic temperature of the gas and
the total pressures employed in these experiments, the vibrational temperatures might depart of the 
kinetic temperature upon the RF excitation.

We used an O$_2$ plasma (0.1 mbar, 50 W) to clean the walls of the cavity which were covered, after the OCS and
CS$_2$ plasma, by a thin layer deposit of a browny material mainly containing sulphur. During the cleaning
process we searched for possible species produced in the cell walls, SO and SO$_2$. 
SO has no transitions within the Q band and therefore only SO$_2$ was searched for. Figure~\ref{so2_cleaning}
shows one of the lines of SO$_2$ observed during the O$_2$ plasma. Before the O$_2$ plasma was initiated, 
the chamber was pumped down to 10$^{-5}$ mbar. Hence, the only sulphur available to
produce SO$_2$ was on the walls. The partial pressure of SO$_2$ from the observed
integrated line intensity was derived to be $\simeq$0.04 mbar.
The mass spectra of the O$_2$ discharge also showed the formation of SO$_2$, whereas the CO signal arising from
possible carbon deposition during the previous OCS and CS$_2$ plasmas was practically negligible.

\subsection{Photodissociation and photochemistry}
\begin{center}
\begin{figure}
\includegraphics[angle=0,scale=0.52]{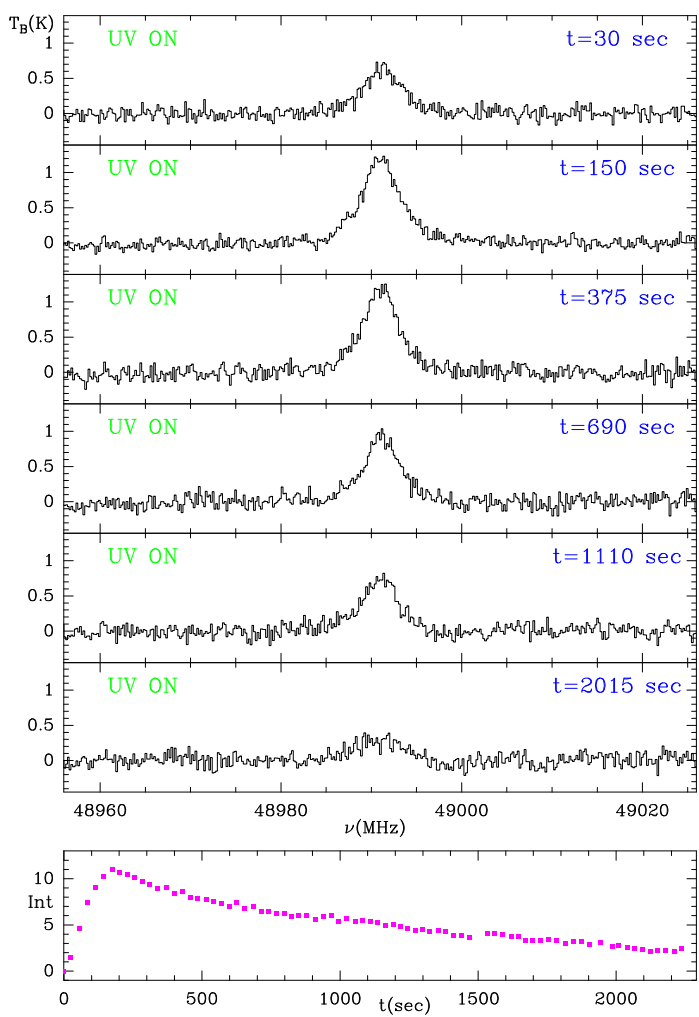}
\caption{Observed $J$=1-0 CS transition at 48990.957\,MHz
using the UV lamp. The integration time was 20 seconds per spectra. The
spectral resolution of the data is 190\,kHz. The upper panels show spectra at selected times while
the lower panel shows the line integrated intensity (K\,$\times$\,MHz) as a function of time. A video showing
the whole set of data is available in the online
version of the paper.}
\label{uv_on}
\end{figure}
\end{center}

The first experiment with the UV lamp was performed with CS$_2$. Figures~\ref{FigSetup}-\ref{Plasma_UV}
show the experimental configuration. 
The gas cell was
filled with CS$_2$ at a pressure of 0.4 mbar. The illumination with the UV lamp was maintained during the
total duration of the experiment with data acquired every 20 seconds with a blanking time
of five seconds imposed by the software of the telescope. A reference spectrum
was performed at
the beginning of the experiment by observing the empty cell during 200 seconds. Hence, the noise
in the data arises essentially from the $ON$ phase of each spectra. The total observing time was 2260
seconds corresponding to 77 spectra (see Figure~\ref{uv_on} and associated video). Several processes
are expected to occur during the experiment. First, as discussed below, CS$_2$ could dissociate into 
CS and S.
Both species can react with CS$_2$ and with the walls of the cell. In turn, CS could be
photodissociated into C and S.

Although wall effects have to be considered in order to study the whole chemistry in the cell,
we have noticed that without UV light the CS$_2$ remains in gas phase as measured with
the mass spectrometer for
times much longer (the longest experiment under these conditions was one hour) than those of the experiment 
with UV light (minutes). 
The data shown in Figure~\ref{uv_on} indicate that
CS reaches its maximum abundance at $\simeq$200 seconds, decreasing smoothly afterwards.
The observed maximum intensity corresponds to a partial pressure of CS, $P$(CS), of $\simeq$10$^{-3}$ mbar.
The line broadening is produced essentially by 
the remaining gas, which, according to mass spectrometric data, is composed mainly of CS$_2$, 
which partially photodissociates, and water desorbed from the walls under UV irradiation, maintaining 
nearly a constant total
pressure, during the experiment.

The gas-phase UV absorption spectroscopy of CS$_2$ has been extensively studied
since 1969 in a wide wavelength range (31-500\,nm) \citep[and references therein]{Keller2013}. Recent high-resolution
absorption measurements in the 105-225\,nm region \citep{Sunanda2015} and in the 205-370\,nm
\citep{Grosch2015} reveal that the spectrum of CS$_2$ presents a highly structured band in the VUV region, 
with a maximum absorption cross section of 3.84$\times$10$^{-16}$ cm$^2$ 
molecule$^{-1}$ located at 198.26\,nm at room temperature \citep{Sunanda2015}. 
For that reason, photodissociation of CS$_2$ is efficient at the 193\,nm
excimer laser wavelength which produces CS($X^1\Sigma^+$, $\nu$ = 1 to 5, $J$ = 5 to 45),
S($^3P$) and S($^1D$) fragments \citep{Yang1980,Butler1980}. \citet{Kitsoupolos2001}
determined the S($^3P_{2,1,0})$/S($^1D_2$) branching ratio to be 1.5$\pm$0.4
from the analysis of the speed distributions of the photofragments. 
In our photochemical studies the UV lamp provides emission lines at 185\,nm and
254\,nm at which absorption cross sections of CS$_2$ are
2.32$\times$10$^{-17}$ cm$^2$ molecule$^{-1}$ \citep{Sunanda2015} and $<10^{-22}$ cm$^2$ molecule$^{-1}$
\citep{Grosch2015}, respectively. Hence, in our experiments CS$_2$ is expected to be almost
exclusively photolyzed by the 185\,nm radiation.

\begin{center}
\begin{figure}
\includegraphics[angle=0,scale=0.52]{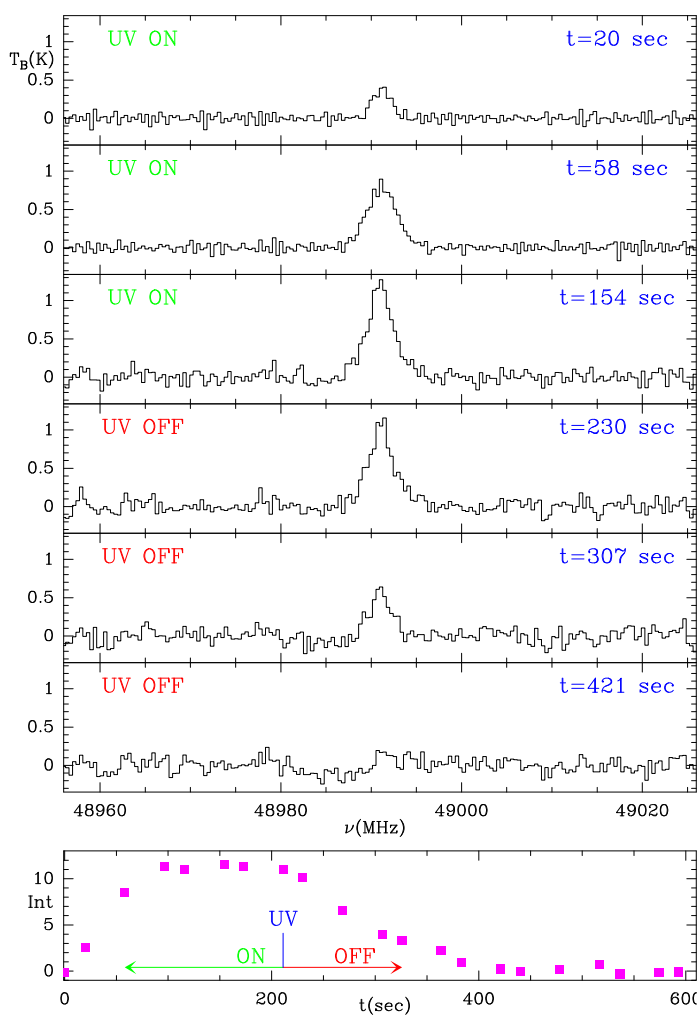}
\caption{Observed $J$=1-0 CS transition at 48990.957\,MHz
using the UV lamp. The integration time was 20 seconds per spectra. The
spectral resolution of the data is 380\,kHz. The upper panels show spectra at selected times while
the lower panel shows the line integrated intensity (in K\,$\times$\,MHz) as a function of time.
The lamp was switched off after 200 seconds. The decrease of the CS abundance is much faster
than in the case of continuous illumination of the cell. A video showing
the whole set of data is available in the online
version of the paper.}
\label{uv_on_off}
\end{figure}
\end{center}

In order to provide some control on the wall effects we repeated the experiment 
but the UV lamp was switched off
after CS reached its maximum (at 200 seconds). Figure~\ref{uv_on_off} shows the
resulting data. At 100 seconds after switching off the lamp, the partial pressure of CS 
measured from the  intensity of the $J$=1-0 line decreased
by a factor of two. After 200 seconds, we could only derive upper limits to $P$(CS).
Hence, a long survival time of CS in the gas phase is observed after its production by photodissociation
is interrupted.

\begin{center}
\begin{figure}
\includegraphics[angle=0,scale=0.52]{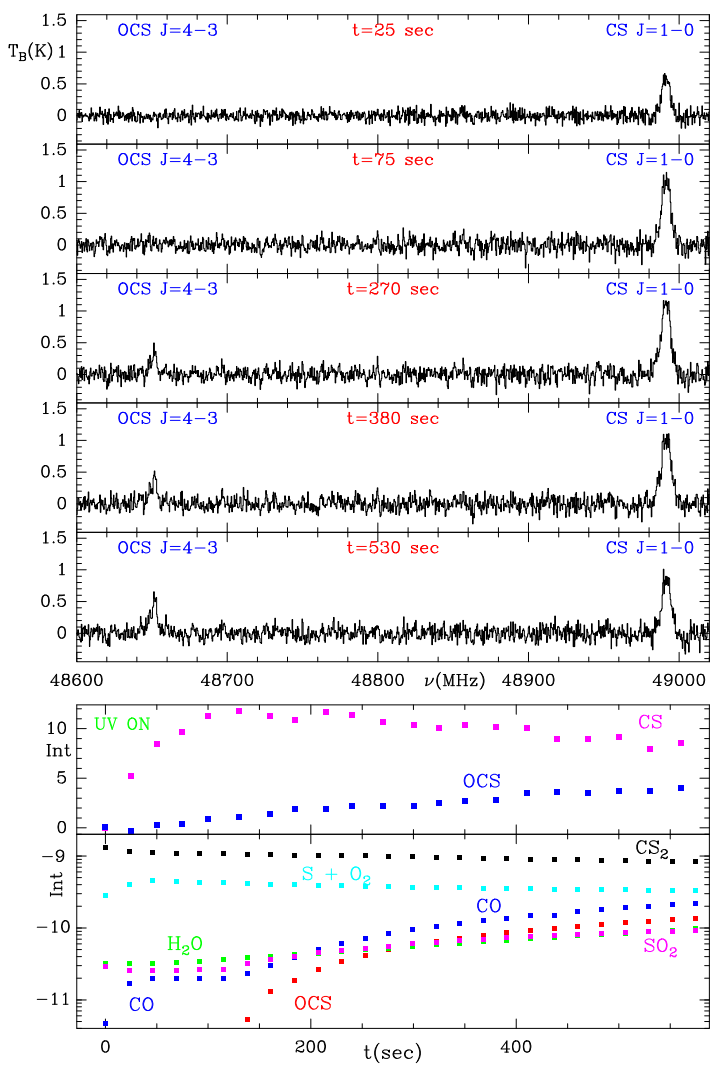}
\caption{Observed transitions CS $J$=1-0 and OCS $J$=4-3 transitions 
using the UV lamp. The integration time was 20 seconds per spectrum. The
spectral resolution of the data is 380\,kHz. The upper panels show spectra at selected times while
the intermediate panel shows the line integrated intensity (in K\,$\times$\,MHz) of CS and OCS as a function of time.
Finally, the bottom panel shows the intensity derived from the mass spectrum for different masses
(CS$_2$, OCS, H$_2$O, S and O$_2$, SO$_2$, and CO).
A video showing the whole set of data is available in the online
version of the paper.}
\label{cs2_o2_uv}
\end{figure}
\end{center}

In a second experiment, we checked the feasibility of chemical reactions in the cell induced by 
photochemistry by introducing 0.4 mbar of CS$_2$ and 0.2 mbar of O$_2$ in the presence of UV
radiation. 
Considering the corresponding
absorption cross sections at 185 nm (2.038$\cdot$10$^{-21}$  cm$^2$ molecule$^{-1}$ for O$_2$
\citep{Yoshino1992} and 2.32$\cdot$10$^{-17}$ cm$^2$ molecule$^{-1}$ for CS$_2$ \citep{Sunanda2015}), CS
radicals are expected to be formed, as well as O atoms but at a lower extend.
The expected molecules to be detected, in addition of CS, are OCS, SO, O$_3$, and SO$_2$.
Unfortunately, as mentioned before, SO does not have rotational transitions in the 
spectral range of the
receiver of the Aries-40m. 
The $J=1-0$ line of CS and the $J=4-3$ line of OCS are separated by 240 MHz and
were observed simultaneously using the same
procedure than in previous cases. Figure~\ref{cs2_o2_uv} shows the results. After CS reaches its
maximum abundance, OCS starts to appear. The partial pressure of CS is again $\simeq$0.001 mbar while that
of OCS is $\simeq$0.002 mbar at times larger than 150 seconds.

The O$_2$ + CS$_2$ reaction has been measured in the
laboratory at temperatures between 1500 and 2100 K and was found to exhibit a very high barrier, 
$\simeq$16000 K \citep{Saito1986}.
Hence, its contribution
to the chemistry in the cell at 300 K is expected to be negligible. 
The reaction of O$_2$ with CS has two product channels, one 
producing OCS + O with a very low rate coefficient ($k$) at ambient temperature
of $k$=2.9$\cdot$10$^{-19}$ cm$^{3}$ s$^{-1}$
\citep{Atkinson1997}, the other 
producing CO + SO with $k$=3.01$\cdot$10$^{-18}$ cm$^{3}$ s$^{-1}$ 
at room temperature. Therefore, formation of OCS from the reaction of CS with molecular 
oxygen is negligible in the gas phase. 
Alternative mechanisms for the production of OCS include wall effects and reaction
of atomic oxygen, product of the photodissociation of O$_2$,
with CS$_2$. 


Atomic oxygen reacts with CS$_2$ at 300 K to produce

\begin{itemize}
\item CS + SO, $k$=3.75$\cdot$10$^{-12}$  cm$^3$\,s$^{-1}$ \citep{Singleton1988}.

\item OCS + S, $k$=3.06$\cdot$10$^{-13}$  cm$^3$\,s$^{-1}$ \citep{Cooper1992}.

\item CO + S$_2$, $k$=1.08$\cdot$10$^{-13}$ cm$^3$\,s$^{-1}$ \citep{Cooper1992}.
\end{itemize}

Contrarily to molecular oxygen, atomic oxygen also reacts relatively fast at ambient temperature with CS:

\begin{equation}
{\rm
O + CS \rightarrow CO + S}
\label{eq:react-04}
,\end{equation}

\noindent
with $k$=2.11$\cdot$10$^{-11}$ cm$^3$\,s$^{-1}$ \citep{Atkinson1997}. Hence, if atomic oxygen
is present with high abundance in the gas phase it will produce OCS, but also distroy
most CS obtained from the photodissociation of CS$_2$. On the other hand, 
the reaction of atomic oxygen with OCS proceeds very slowly, $k$=1.36$\cdot$10$^{-14}$ cm$^3$\,s$^{-1}$
\citep{Singleton1988}. Hence OCS, if formed from the reaction of O and CS$_2$, could survive in
the gas phase, being only destroyed by wall reactions and by UV photons.

The mass spectra
obtained during the experiment help to understand the chemical processes occurring in the cell.
The main observed species are CS$_2$ (76), SO$_2$ (64), OCS (60), S/O$_2$ (32), CO (28), H$_2$O (18).
At the beginning, a fast increase of mass peak 32 is observed due to the photodissociation
of CS$_2$ which overcompensates the photodissociation of O$_2$. Then, both CS$_2$ and mass peak 32 
decrease by a factor of two during the experiment. We note that OCS, as
traced by the emission of its
$J$=4-3 line, requires nearly 150 seconds to reach an abundance high enough to be detected with 
the receivers. A similar time is also required to produce a significant signal in the mass 
spectrometer for OCS.
The first sharp increase of CO is associated to reaction
\ref{eq:react-04}. Later, once OCS is formed then CO can be produced by the photodissociation of OCS.
The threshold for photodissociation of OCS is near 255\,nm \citep{Lochte1932} and the photofragments
produced are CO, in its electronic ground state, and sulphur atoms in its $^1D_2$ state.
In our experiment the photodissociation is dominated by the absorption of the 185\,nm line which
has a cross section of $\simeq$3$\cdot$10$^{-19}$ cm$^2$ molecule$^{-1}$ 
\citep{Keller2013}.

The mass peak corresponding to H$_2$O
(which is certainly desorbed from the walls) shows a flat distribution following a very similar
time dependence to that of SO$_2$ which is probably formed on the walls, since the
gas phase reaction

\begin{equation}
{\rm
SO + O_2 \rightarrow SO_2 + O}
\label{eq:react-05}
,\end{equation}

\noindent
proceeds very slowly at ambient temperature, $k\simeq$8$\cdot$10$^{-17}$\,cm$^3$ s$^{-1}$
\citep{Atkinson1997,DeMore1997}. 


O$_3$ was not observed in our experiment. However, in the Schumann-Runge band spectral region 
(175-200\,nm), O$_2$ can be photolyzed forming
O atoms via these two dissociative channels:

\begin{equation}
{\rm
O_2(X^3\Sigma^-) + h\nu (\lambda<\,240\,nm) \rightarrow O(^3P) + O(^1D)}
\label{eq:react-03}
\end{equation}
\noindent
and O($^3P$) + O($^3P$).
At $\lambda<$ 177\,nm, the quantum yield of O($^1D$) is unity, while it is zero at longer wavelengths
\citep{Nee1997}. Therefore, oxygen atoms are formed in the ground state,$^3P$, when O$_2$ is
photodissociated at $\lambda$=185\,nm.

O($^3P$) atoms can react rapidly with O$_2$ in a three-body reaction 
to form ozone, O$_3$:
\begin{equation}
{\rm O(^3P)+O_2+M\rightarrow O_3+M}
,\end{equation}

\noindent
where M is a third body (O$_2$ or CS$_2$). O$_3$ is photodissociated efficiently by the
radiation at 254\,nm of the mercury lamp:
\begin{equation}
{\rm O_3 + h\nu(254 nm) \rightarrow O(^1D)+O_2}
,\end{equation}
\noindent
followed by
\begin{equation}
{\rm O(^1D)+M\rightarrow O(^3P)+M }
.\end{equation}

We can therefore understand why O$_3$ was not observed in our cell. First,
the three body reaction producing it proceeds slowly at our low
experimental pressures. In addition,
the lamp flux at 254 nm is about five times that at 185 nm, and the absorption
cross section of O$_3$ at 254 nm is 1.157$\cdot10^{-17}$
cm$^2$molec$^{-1}$ \citep{Molina86}. Consequently,
ozone concentration in the mixture is expected to be very small
and, furthemore, the intensity of the lines is rather low (I(300 K)= 9.72$\times$10$^{-7}$ 
nm$^2$\,MHz
for the $1_{1,1}-2_{0,2}$, the strongest line in this region).
Hence, the expected intensities for these ozone lines are below our detection limit in the present experiments.

A detailed analysis of the different involved chemical processes would require additional 
experiments at different
partial pressures of CS$_2$ and O$_2$, and different intensities of the UV lamp. 
Nevertheless, the
three experiments shown in this section demonstrates the capacity of radio astronomical receivers
to perform chemical experiments of astrophysical interest. 

\section{Conclusions}
In this work we have shown that radio astronomical observation techniques and receivers 
can be used in laboratory experiments to perform high-sensitivity molecular spectroscopy of molecules 
inside a gas cell at low pressure.
Compared with other techniques at low frequency the sensitivity offered by radio astronomical
receivers is not as performant and will require longer integration times. However, even at
these frequencies, the very large instantaneous bandwidth (around 20 GHz), spectral purity, 
and the linear dependence of line intensity with partial pressures, provides an alternate way 
to perform experiments of interest in the field of astrochemistry and laboratory spectroscopy. 
At millimeter and submillimeter wavelengths, with an instantaneous bandwidth of 40 GHz and a high 
spectral purity, emission spectroscopy becomes comparable in sensitivity with narrow band
laboratory absorption techniques and permits to follow the time variation of many species
simultaneously in chambers devoted to laboratory simulations of the chemistry of interstellar
clouds.

The selection of appropriated windows for the cell
has been addressed. 
Standard quartz windows, which would be a natural choice in ultra-high vacuum conditions,
are not adapted to reach the sensitivity
provided by the heterodyne receivers. 
The interference between the two faces of the quartz window {worsens}
performance and appropriately designed anti-reflection coatings are important.
If total pressures above 10$^{-4}$ mbar
prevail in the cell, then Upilex windows can
be used providing a very good transmission and a minimum degradation of the
receiver performance.  Nevertheless, other materials with appropriate antireflection
treatments will be tested. First results using Teflon windows indicate that this
material offers very good transmission and safety conditions.

We showed that molecular spectroscopy can be performed in the laboratory  by using radio astronomical receivers and spectrometers, 
which are coupled with a small size gas cell placed filled with a static (stable molecules) or flowing 
gas (for unstable species).
Standard ON/OFF or FSW observing modes can be performed depending on the type and goals of
the experiments. Special care has to be taken with the stability of the cold load.

Radio frequency cold plasmas can be produced in the gas cell without any perturbation
of the receivers and their associated electronics. We have generated cold plasmas 
of different molecular species and successfully detected partial pressures of their 
products as low as 10$^{-3}$ mbar.

UV photochemistry can be studied in the cell. 
Partial pressures of photochemical
products as low as 10$^{-3}$ mbar have been observed in our experiments. No perturbation has been found between
the lamp discharge and the receiver electronics.

The residence time in the gas phase of reactive species produced in plasma or
the photolysis experiments, such as CS, is large enough to allow further chemical reactions
with potential in studying the chemistry of photodissociation regions in the ISM.
To conclude, we expect that the gas cell that will be soon operational for the Nanocosmos project, 1\,m long,
with the 31.5-50 and 72-116.5 GHz frequency domain covered by
HEMT receivers, coupled to 40 GHz wide Fast Fourier Transform spectrometers,
will open new perspectives in laboratory astrophysics.

\begin{acknowledgements}
The research leading to these results has received funding from
the European Research Council under the European Union's Seventh Framework Programme 
(FP/2007-2013) / ERC-SyG-2013 Grant Agreement n. 
610256 NANOCOSMOS
and from spanish MINECO CSD2009-00038 (ASTROMOL) under 
the Consolider-Ingenio Program. We also thank spanish MINECO for funding under grants 
AYA2012-32032, AYA2016-75066-C2-1-P, FIS2013-48087-C2-1-P, FIS2016-77726-C3-1-P, 
FIS2016-77578-R, MAT2014-54231-C4-1-P. 
\end{acknowledgements}

\begin{appendix}
\section{Line profiles}
The effect of the lineshape on the
observed spectrum deserves some consideration because, for a given integrated
line intensity, it will determine the peak value of the detected
signal, and therefore the attainable signal to noise ratio for a given integration time.

The main line broadening mechanisms at the frequencies covered by the receiver
and at the typical pressure and temperature operating conditions are thermal
(Doppler) broadening and collisional (pressure) broadening. Neglecting more
subtle effects such as Dicke narrowing, the speed dependence of the relaxation
rates or line-mixing effects \citep{har08}, the lineshape functions describing the Doppler and
the collisional broadening are the Gaussian function and the Van Vleck-Weisskopf
function, respectively. The normalized Doppler lineshape function can be
expressed as
\begin{equation}
\phi_D(\nu)=\frac{1}{\Delta\nu_D}\sqrt{\frac{\ln(2)}{\pi}}e^{-\ln(2)((\nu-\nu_0)/\Delta\nu_D)^2},
\label{eqn:Gaussnorm}
\end{equation}
\noindent with
\begin{equation}
\Delta\nu_D=3.581\cdot10^{-7}\nu_0\sqrt{\frac{T}{M}},
\end{equation}
\noindent where $\Delta\nu_D$ is the Gaussian half width at half maximum, $T$
is the temperature in K and $M$ the molecular mass in unified atomic mass units.
As an example, at 300 K, and at 45 GHz, the Doppler full width at half maximum
(FWHM) of a molecule of M=50 amu is $2\Delta\nu_D=79$ kHz.

The normalized Van Vleck-Weisskopf lineshape profile can be written
\begin{equation}
\phi_{VVW}(\nu)=\frac{\nu}{\pi\nu_0}\left(\frac{\Delta\nu}{(\Delta\nu)^2 +
(\nu-\nu_0)^2}+\frac{\Delta\nu}{(\Delta\nu)^2 + (\nu+\nu_0)^2}\right)
,\end{equation}
with $\Delta\nu=(2\pi t_2)^{-1}$, and $t_2$ the average time
between molecular collisions. When $\nu_0\gg\Delta\nu$, the Van Vleck-Weisskopf
profile reduces to the Lorentz lineshape with
$\Delta\nu=\Delta\nu_C$, the HWHM of the Lorentz profile, that is

\begin{equation}
\phi_L(\nu)=\frac{1}{\pi}\frac{\Delta\nu_C}{(\nu-\nu_0)^2+\Delta\nu_C^2}
\label{eqn:Lorentznorm}
\end{equation}

The collisional width in this model increases linearly with pressure (or
density), i.e.: $\Delta\nu_C=\Delta\nu_C^0\times P$, where $\Delta\nu_C^0$ is
the pressure broadening coefficient
normally expressed in units of
MHz/mbar, MHz/Torr, or \wn/atm) and $P$ is the pressure. Note that pressure broadening
coefficients in the literature most frequently refer to the Lorentz halfwidth at
half maximum. They depend on the quantum numbers of the transition, the
temperature and the colliding partner, but typical values are 2--5 MHz/mbar. In a
mixture of species the collisional width is additive, that is:

\begin{equation}
\Delta\nu_C=P\sum_{A}X_A(\Delta\nu_C^0)_A
,\end{equation}

\noindent where $P$ is the total pressure, $X_A$
is the mole fraction of species $A$ and $(\Delta\nu_C^0)_A$ its corresponding
pressure broadening coefficient. As an example, for pressures 1--0.01 mbar and a
typical $\Delta\nu_C^0=3$ MHz/mbar, $\Delta\nu_C$=3--0.03 MHz. It can be seen
that, except for rather low pressures, the dominant broadening mechanism in the
cell will be collisional broadening, but with a non-negligible contribution of
Doppler broadening. In this regime, and, again, neglecting speed-dependent and
other effects, the Voigt profile renders a good representation of the spectral
shape.  Moreover, for the highest pressures where the cell will be operated
(a few mbar), the Lorentz lineshape can be used.

Mathematically, the Voigt profile is the convolution of a Gaussian and a
Lorentzian function, and, physically, implies that both broadening
mechanisms are totally uncorrelated, i.e., that each velocity class has its own
spectral distribution owing to collisions. The Voigt profile has no analytical
expression, but it can be written as
\begin{equation}
\phi_V(\nu)=\sqrt{\frac{\ln(2)}{\pi}}\frac{1}{\Delta\nu_D}\frac{a}{\pi}
\int_{-\infty}^{+\infty}\frac{e^{-y^2}}{a^2+(w-y)^2}dy,
\end{equation}
with $w=\sqrt{\ln(2)}(\nu-\nu_0)/\Delta\nu_D$ and
$a=\sqrt{\ln(2)}\Delta\nu_C/\Delta\nu_D$. Combining $a$ and $w$ into the complex
variable $z=a+iw$, the Voigt function can be represented in terms of the complex
error function $W(z)$:
\begin{equation}
\phi_V(z)=\sqrt{\frac{\ln(2)}{\pi}}\frac{1}{\Delta\nu_D}\Re[W(z)]
\end{equation} which
can be computed efficiently by numerous algorithms. In the low pressure limit,
where $\Delta\nu_D\gg\Delta\nu_C$ ($a\rightarrow 0$), the Voigt profile renders
the Doppler profile, while at higher pressures, when
$\Delta\nu_C\gg\Delta\nu_D$ ($a\gg 1$), it reduces to the Lorentz
profile.

As can be inferred from equations~\ref{eqn:Gaussnorm} and \ref{eqn:Lorentznorm},
in the Doppler limit the
intensity at the line centre will be proportional to pressure (or, equivalently,
to number density), and the linewidth
will remain constant.  On the contrary, in the Lorentz limit, the intensity at
the line centre will remain constant, and the linewidth will increase linearly
with pressure.
From a practical standpoint, there are two main consequences for the experiments
described here:  on the one hand, increasing the partial pressure of a species beyond
a certain value will not increase the signal intensity, although the larger width
may allow for a higher degree of smoothing, thereby rendering a higher signal
to noise ratio.  On the other hand, in a mixture of gases, the peak signal of
a minor component may be decreased by the broadening caused by a more abundant species.

\end{appendix}
\end{document}